\documentclass[]{mn2e}
\usepackage{epsfig}
\def\lsim{\vcenter{\hbox{$<$}\offinterlineskip\hbox{$\sim$}}}
\def\gsim{\vcenter{\hbox{$>$}\offinterlineskip\hbox{$\sim$}}}
\title[Infrared stellar populations in the central parts of the Milky Way
galaxy]{Infrared stellar populations in the central parts of the Milky Way
galaxy\thanks{This is paper no.\ 17 in a refereed journal based on data from
the ISOGAL project.}}
\author[Jacco Th. van Loon et al.]
{Jacco Th. van Loon$^1$,
 G.F. Gilmore$^2$,
 A. Omont$^3$,
\newauthor
 J.A.D.L. Blommaert$^4$,
 I.S. Glass$^5$,
 M. Messineo$^6$,
\newauthor
 F. Schuller$^3$,
 M. Schultheis$^3$,
 I. Yamamura$^7$
 and H.S. Zhao$^2$\\
$^1$Astrophysics Group, School of Chemistry \& Physics, Keele University,
    Staffordshire ST5 5BG, United Kingdom\\
$^2$Institut d'Astrophysique de Paris, CNRS, 98bis Boulevard Arago, F-75014
    Paris, France\\
$^3$Institute of Astronomy, Madingley Road, Cambridge CB3 0HA, United
    Kingdom\\
$^4$Instituut voor Sterrenkunde, Katholieke Universiteit Leuven,
    Celestijnenlaan 200 B, B-3001 Leuven, Belgium\\
$^5$South African Astronomical Observatory, P.O.Box 9, Observatory 7935, South
    Africa\\
$^6$Leiden Observatory, P.O.Box 9513, NL-2300 RA Leiden, The Netherlands\\
$^7$Institute of Space and Astronautical Science, 3-1-1 Yoshinodai,
    Sagamihara, Kanagawa 229, Japan}
\date{Accepted ????.
      Received ????;
      in original form ????}
\pagerange{\pageref{firstpage}--\pageref{lastpage}}
\pubyear{????}
\begin{document}
\maketitle
\label{firstpage}
\begin{abstract}
Near- and mid-IR survey data from DENIS and ISOGAL are used to investigate the
structure and formation history of the inner $10^\circ$ (1.4 kpc) of the Milky
Way galaxy. Synthetic bolometric corrections and extinction coefficients in
the near- and mid-IR are derived for stars of different spectral types, to
allow the transformation of theoretical isochrones into observable
colour-magnitude diagrams. The observed IR colour-magnitude diagrams are used
to derive the extinction, metallicity and age for individual stars. The inner
galaxy is dominated by an old population ($\gsim7$ Gyr). In addition, an
intermediate-age population ($\sim200$ Myr to 7 Gyr) is detected, which is
consistent with the presence of a few hundred Asymptotic Giant Branch stars
with heavy mass loss. Furthermore, young stars ($\lsim200$ Myr) are found
across the inner Bulge. The metallicities of these stellar population
components are discussed. These results can be interpreted in terms of an
early epoch of intense star formation and chemical enrichment which shaped the
bulk of the Bulge and nucleus, and a more continuous star formation history
which gradually shaped the disk from the accretion of sub-solar metallicity
gas from the halo. A possible increase in star formation $\sim200$ Myr ago
might have been triggered by a minor merger. Ever since the formation of the
first stars, mechanisms have been at play that mix the populations from the
nucleus, Bulge and disk. Luminosity functions across the inner galactic plane
indicate the presence of an inclined (bar) structure at ${\gsim}1$ kpc from
the galactic centre, near the inner Lindblad resonance. The innermost part of
the Bulge, within $\sim1$ kpc from the galactic centre, seems azimuthally
symmetric.
\end{abstract}


\section{Introduction}

\subsection{The central parts of the Milky Way galaxy}

The Milky Way galaxy provides a unique opportunity to learn about the
formation, the structure and the evolution of galaxies. The central parts of
the galactic bulge and disk have remained elusive, though, due to the
extremely high extinction at short wavelengths and poor spatial resolution at
longer wavelengths. Most of the current belief that the stellar content of the
galactic bulge is old, $t\gsim10$ Gyr, and metal-rich, $[M/H]\sim$solar,
results from studies in low extinction regions (e.g.\ Baade's Window) at
galacto-centric radii $R>500$ pc (Rich 1998a). With the advent of infrared
(IR) cameras and adaptive optics techniques, the exploration of the galactic
centre has revealed the presence of massive stars that indicate recent star
formation (Genzel et al.\ 1994). Yet data concerning the relationship between
the central parsec of the galaxy and the Bulge, halo and disk remains scarce.

What we observe today is the time-integrated history of star formation, gas
flows, and mergers in the galaxy, and it may be envisaged that the formation
of the different components of the galaxy --- halo, Bulge, nucleus and disk
--- are not independent events. Common preconceptions about the Bulge being an
old, metal-rich small elliptical galaxy are being challenged (Wyse et al.\
1997). For instance, recent near-IR photometry and spectroscopy of stars in
the inner Bulge ($R\lsim500$ pc) suggest the presence of an intermediate-age
population ($t\sim1$ to 2 Gyr: e.g.\ Frogel 1999a). Did these stars form in
the Bulge, or in the nucleus? Is there any connection between the star
formation history and the formation of galactic structures such as a bar
(Blitz et al.\ 1993) or tri-axial Bulge (Nakada et al.\ 1991)?

The galactic bulge is fundamentally typical of all bulges in late-type spirals
(Frogel 1990). In particular, it is very similar to that of M31 and M32
(Davies et al.\ 1991; DePoy et al.\ 1993; Davidge 2000b, 2001; Rich 2001;
Stephens et al.\ 2001), and the central nuclei of M33 (Mighell \& Rich 1995;
Davidge 2000a; Mighell \& Corder 2002; Stephens \& Frogel 2002) and NGC 247
and NGC 2403 (Davidge \& Courteau 2002), which all seem predominantly old and
metal-rich but most of which do contain bright AGB stars and possibly even
younger populations. Rich \& Mighell (1995) ask the question why the
integrated light of the bulge of M31 is so red despite the presence of an
intermediate population. This is probably due to the fact that the integrated
light results mainly from the red giants of $\sim$solar mass, and in addition
from the many red dwarfs that have been formed in any generation of stars.
This explains why observations of distant bulges show an old, metal-rich
content, whereas observations of spatially resolved stellar populations in
nearby bulges increasingly show the presence of younger as well as
metal-poorer stellar populations --- see also Lamers et al.\ (2002) for the
bulge of M51 and Rejkuba et al.\ (2001) for the giant elliptical NGC5128.

\subsection{The ISOGAL project}

We successfully completed a massive survey of the inner galaxy with ISO at 7
and 15 $\mu$m (ISOGAL) to study galactic structure and astrophysics (Omont et
al.\ 2002; Schuller et al.\ 2002). Within $\sim1^\circ$ from the galactic
plane, surveys at wavelengths $\lambda\lsim1$ $\mu$m are dominated by
foreground objects in the galactic disk. Surveys at mid-IR wavelengths
($\lambda\gsim5$ $\mu$m), however, are dominated by luminous stars in the
inner galaxy, at a distance of $d\sim8$ kpc. Indeed, cross-correlation of the
ISOGAL survey with sources from the DENIS survey (I, J and K$_{\rm s}$-bands)
within galactic longitudes $|l_{\rm II}|<6^\circ$ and latitudes $|b_{\rm
II}|<4^\circ$ initially resulted in the identification of $\sim9000$ stars
detected at 2.2 and 7 $\mu$m, mainly M-type stars on the first and asymptotic
giant branches (RGB and AGB, respectively) within the inner kpc of the galaxy.
First results have been published by Omont et al.\ (1999), Glass et al.\
(1999), Schultheis et al.\ (2000), Alard et al.\ (2001) and Ojha et al.\
(2002).

%
%
\begin{figure}
\centerline{\psfig{figure=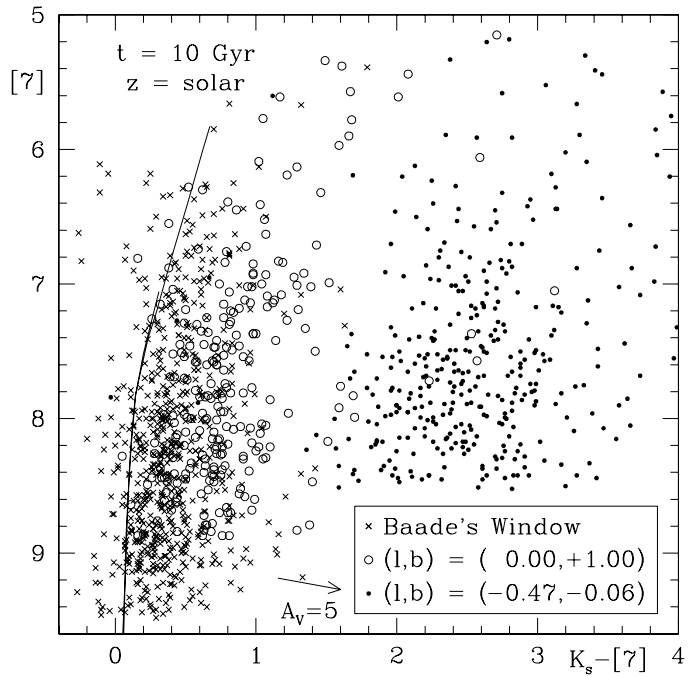,width=90mm}}
\caption[]{Location in the $[7]$ versus $(K_{\rm s}-[7])$ diagram of stars in
three fields in the inner galaxy that suffer from different amounts of
interstellar extinction. An isochrone for an unreddened 10 Gyr old population
of solar metallicity is plotted for reference.}
\end{figure}

An example of the combined ISOGAL/DENIS data is presented in Fig.\ 1: the
stellar populations in three fields in the inner galactic bulge are clearly
displaced in the $[7]$ versus $(K_{\rm s}-[7])$ colour-magnitude diagram by
different amounts of interstellar extinction. Differences in age, metallicity
and distance affect the location of stars in the colour-magnitude diagram too.
For instance, the stars in the direction of $(l_{\rm II},b_{\rm
II})=(-0.47,-0.06)$ cannot be described simply by a single population of
uniform age and metallicity at a fixed distance behind a veil of uniform
extinction, as evidenced by the non-uniform spread in the $(K_{\rm s}-[7])$
colours. Here we use the IR photometry to determine these properties of the
stellar populations, as well as the structure of the inner galaxy (see also
Frogel et al.\ 1999).

The techniques for deriving the ages, metallicities and extinctions for
individual stars from infrared photometry are developed and described in
Sections 3 and 4. Distributions of age and metallicity for the inner
$10^\circ$ (1.4 kpc) of the galaxy are presented in Sections 5 and 6.
The extinction-corrected luminosity funtions are used to probe differential
depths. The discussion (Section 7) addresses the history of the star formation
and chemical enrichment, the distinction between Bulge, disk and nucleus, and
the presence of a bar.

\section{Infrared data}

%
%
\begin{figure}
\centerline{\psfig{figure=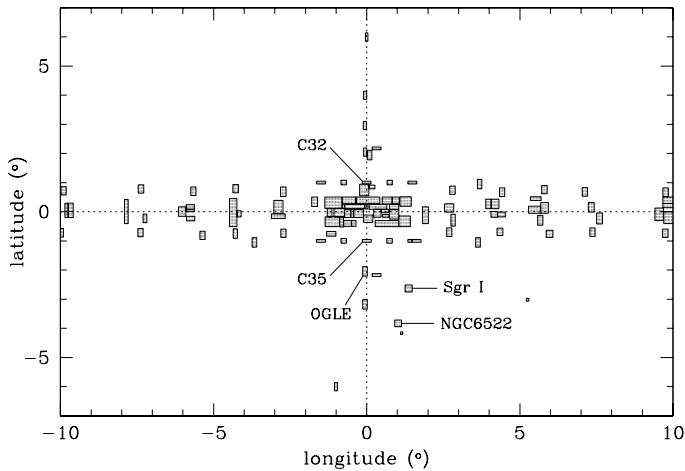,width=90mm}}
\caption[]{Location in the galaxy of the ISOGAL fields used here.}
\end{figure}

From the first version of the ISOGAL/DENIS point source catalogue (PSC1.0),
all fields within $10^\circ$ of the galactic centre have been selected (Fig.\
2). The data comprises mid-IR imaging photometry at 7 and 15 $\mu$m obtained
with the ISO-CAM instrument onboard ISO for the ISOGAL project (Omont et al.\
2002), and near-IR imaging photometry in the I, J and K$_{\rm s}$-bands
obtained at ESO La Silla within the framework of the DENIS survey (Epchtein et
al.\ 1999; Simon et al.\ in preparation). Details of the photometry and the
catalogue can be found in Schuller et al.\ (2002). We should mention that the
limiting magnitudes applied to the selection of 7 \& 15 $\mu$m sources vary
from field to field according to the density of sources and the background
level. These are generally about a magnitude brighter than for the preliminary
data, but as a result the photometry from the PSC1.0 is believed to be more
reliable and homogeneous, with a new calibration yielding mid-IR magnitudes
brighter by 0.4 mag on average. This is very important when comparing the
photometry obtained in different fields.

Because different ISOGAL fields have been observed in different observing
modes (combination of ISO-CAM filter and pixel size), the following
transformations have been made to standardize the 7 and 15 $\mu$m photometry
to the LW2 and LW3 filters at a pixel size of $6^{\prime\prime}$ (Appendix A):
\begin{displaymath}
\left[7\right],\left[15\right] = \hspace{-1mm} \left[ \begin{array}{l}
\left[LW2\hspace{2mm} 3^{\prime\prime}\right] -0.002 \\
\left[LW2\hspace{2mm} 6^{\prime\prime}\right]        \\
\left[LW5\hspace{2mm} 3^{\prime\prime}\right] +0.122 \\
\left[LW5\hspace{2mm} 6^{\prime\prime}\right] +0.124 \\
\left[LW6\hspace{2mm} 3^{\prime\prime}\right] +0.064 \\
\left[LW6\hspace{2mm} 6^{\prime\prime}\right]+0.066 
\end{array} \right] ,
\left[ \begin{array}{l}
\left[LW3\hspace{2mm} 3^{\prime\prime}\right] -0.093 \\
\left[LW3\hspace{2mm} 6^{\prime\prime}\right]        \\
\left[LW9\hspace{2mm} 3^{\prime\prime}\right] -0.052 \\
\left[LW9\hspace{2mm} 6^{\prime\prime}\right] +0.131
\end{array} \right]
\end{displaymath}

\section{From Hertzsprung-Russell Diagram to colour-colour and
colour-magnitude diagrams}

\subsection{Bolometric corrections and colours}

To compare IR photometry with theoretical isochrones, transformations are made
from the physical properties of stars (luminosity $L$, effective temperature
$T_{\rm eff}$ and metallicity $z$; the gravity $g$ is a $(L,T_{\rm eff})$
dependent parameter) to the observable properties of stars (magnitudes and
colours).

%
%
\begin{figure*}
\centerline{\psfig{figure=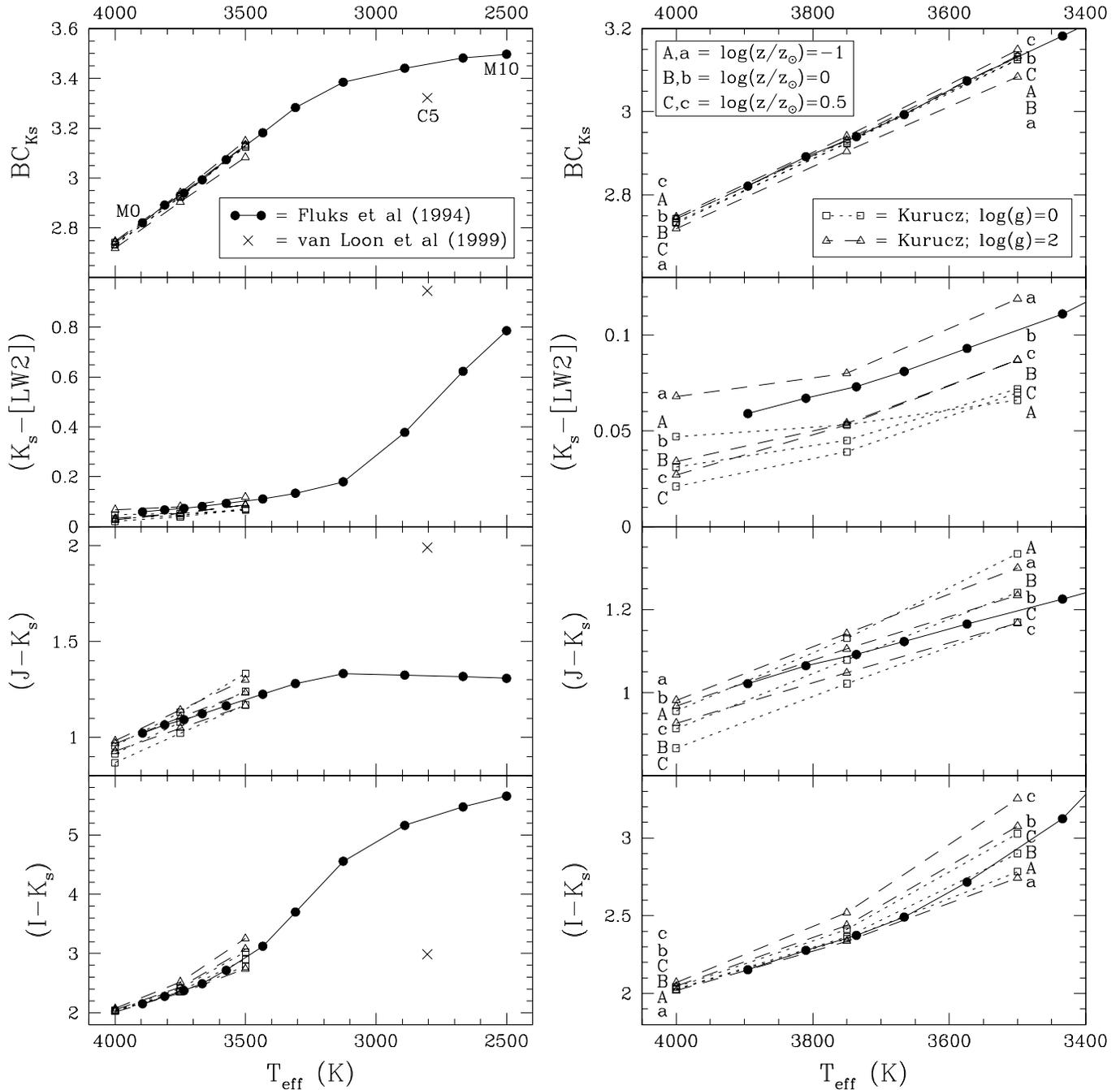,width=180mm}}
\caption[]{Bolometric correction to the K$_{\rm s}$-band and IR colours for
synthetic (dust-free) M-type spectra from Kurucz (1993; squares for gravity
$\log{(g[{\rm cm/s^2}])}=0$ and triangles for $\log{(g)}=2$) and Fluks et al.\
(1994; dots), and for a carbon star (van Loon et al.\ 1999; cross).}
\end{figure*}

Bolometric corrections are computed for each filter band by (i) convolving a
Kurucz (1993) model spectrum with the filter transmission curve, and (ii)
comparing this to the result obtained when using a model spectrum appropriate
for an A0 V star: a ``Vega'' model with $T_{\rm eff}=9550$ K,
$\log(z/z_\odot)=-0.5$, and $\log(g)=3.95$ (Kurucz 1993). However, very cool
stars with $T_{\rm eff}\lsim3500$ K exhibit strong photospheric absorption by
molecules, and this is not satisfactorily incorporated within the Kurucz
(1993) models. For these very cool stars Fluks et al.\ (1994) used the MARCS
code to compute synthetic spectra for M-type giants down to $T_{\rm eff}=2500$
K, and these are used to extrapolate the Kurucz results to the coolest
spectral types (assuming no circumstellar dust).

The Kurucz models indicate dependencies of magnitudes and colours on
metallicity and gravity (Fig.\ 3). The bolometric correction to the
K$_{\rm s}$-band is rather insensitive to metallicity and gravity except
perhaps for very cool, metal-poor RGB stars with $\log(z/z_\odot)=-1$ and
$\log(g)=2$. The $(K_{\rm s}-[LW2])$ colour shows a complicated dependence on
metallicity and gravity, but the differences are only a few 0.01 mag. The
$(J-K_{\rm s})$ colour depends more on metallicity than on gravity. Due to the
strong absorption by TiO molecules within the I-band the $(I-K_{\rm s})$
colour becomes redder when metallicity increases, but also when gravity
increases for $\log(z/z_\odot)>-1$: the RGB stars have redder $(I-K_{\rm s})$
colours than AGB stars of the same $T_{\rm eff}$.

The Fluks et al.\ models represent somewhat metal-poor RGB stars. For very
cool stars with $T_{\rm eff}<3500$ K, a small offset is applied to the Fluks
et al.\ models such that the bolometric correction for their M5 III model
coincides with an extrapolation from the Kurucz models down to the
corresponding $T_{\rm eff}=3434$ K. This combines the information for very
cool stars from Fluks et al.'s MARCS models with the metallicity and gravity
grid from Kurucz's ATLAS models.

These models do not apply to carbon stars --- AGB stars whose photospheric
chemistry has switched from oxygen-dominated to carbon-dominated. As an
illustration for the expected bolometric correction and colours for a carbon
star, the template spectral energy distribution for AQ Sgr ($T_{\rm eff}=2804$
K) from van Loon et al.\ (1999) is used (Fig.\ 3). Carbon stars are very red
in $(J-K_{\rm s})$ while relatively blue in $(I-K_{\rm s})$. In the remainder
of this paper, we assume that there are no carbon stars in the ISOGAL sample
towards the inner galactic bulge (Frogel et al.\ 1990; Tyson \& Rich 1991).

Our procedure is largely analogous to that of Girardi et al.\ (2002). Alvarez
et al.\ (2000) also studied the near-IR bolometric corrections and the
strength of molecular absorption in the spectra of late-type giants. Their
results are very similar to ours, which can be seen from the good match of the
I-band bolometric correction as a function of $(I-J)$ colour (their Fig.\ 7).
Bessell et al.\ (1998) studied the bolometric corrections and effective
temperature scales of stars including late-type giants, using a more recent
version of the MARCS model atmospheres. Again, there is a close match with our
results (compare with their Figs.\ 16--19), but our K$_{\rm s}$-band
bolometric corrections are larger by $\sim0.1$ mag --- something which may be
ascribed to the different K$_{\rm s}$-band filters that are referred to.
Unfortunately, their work does not deal with very cool giants ($T_{\rm
eff}\lsim3500$ K).

\subsection{Mid-IR absorption by photospheric molecules}

It was suspected by Glass et al.\ (1999) that the negative colours $(K_{\rm
s}-[LW2])\sim-0.3$ mag of RGB stars seen in the preliminary ISOGAL/DENIS data
of Baade's Window are due to photospheric absorption by molecules in the
spectral region around 7 $\mu$m. Although the PSC1.0 photometry no longer
shows such negative colours (Fig.\ 1), it is still worth investigating the
presence of molecular absorption because it might affect the comparison
between photometry and stellar models. The transformation between the LW2 and
LW5 filters (Section 2 \& Appendix A), for instance, is affected by absorption
by SiO and H$_2$O molecules in the LW2 band whilst the LW5 band is virtually
free of molecular absorption (Aringer et al.\ 1997; J{\o}rgensen et al.\
2001; Tsuji 2001).

%
%
\begin{figure}
\centerline{\psfig{figure=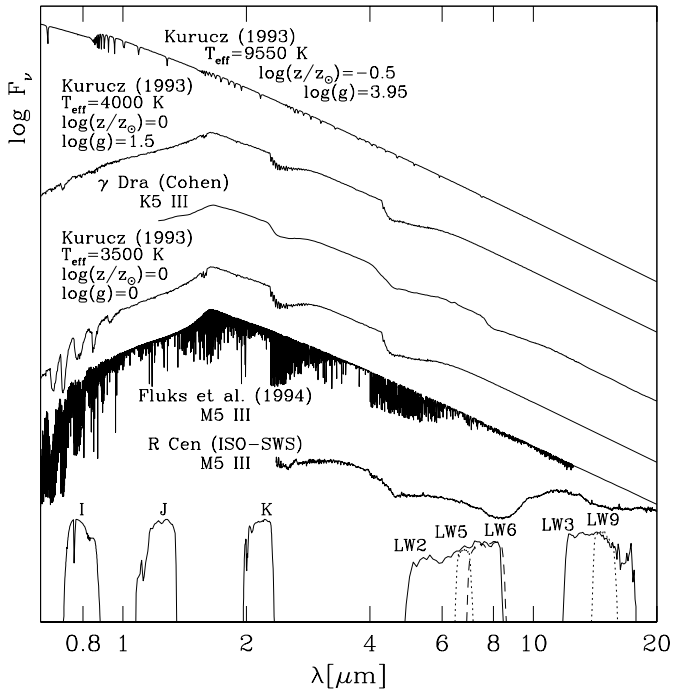,width=88mm}}
\caption[]{Synthetic and observed spectra for Vega and K5 and M5 giants,
illustrating the prominent molecular absorption at IR wavelengths. The
transmission curves of the DENIS and ISO-CAM filters are also plotted.}
\end{figure}

Several model and observed IR spectra are displayed in Fig.\ 4, together with
the DENIS and ISO-CAM filter transmission curves. The observed spectrum of the
K5 giant $\gamma$ Dra (Cohen et al.\ 1999) is compared with a $T_{\rm
eff}=4000$ K, $\log(z/z_\odot)=0$, $\log(g)=1.5$ Kurucz model spectrum, and
the observed ISO-SWS spectrum of the M5 giant R Cen (Yamamura \& de Jong 2000)
is compared with a corresponding Fluks et al.\ model spectrum and with a
$T_{\rm eff}=3500$ K, $\log(z/z_\odot)=0$, $\log(g)=0$ Kurucz model spectrum.
Absorption around 5 $\mu$m is produced by both the Kurucz model spectra (due
to CO with a band head at $\lambda\sim4.3$ $\mu$m) and the Fluks et al.\ model
spectra (due to CO, and SiO with a band head at $\lambda\sim4.0$ $\mu$m), and
is consistent with each other and with the observed spectra despite the fact
that these models do not include absorption by H$_2$O at $\lambda\sim6$
$\mu$m. Some absorption around 8 $\mu$m is produced by the Fluks et al.\
models (but not by the Kurucz models), but these spectra do not reproduce the
observed absorption in the SiO fundamental band (with a band head at
$\lambda\sim7.6$ $\mu$m) in the spectrum of R Cen. At $\lambda{\gsim}8$
$\mu$m, the spectrum of R Cen is greatly affected by circumstellar dust
emission.

Is there agreement between the observed IR colours and the IR colours deduced
from the model spectra? The $(K_{\rm s}-[LW2])$ and $([LW2]-[LW5])$ colours of
$\gamma$ Dra are 0.06 and 0.08 mag, respectively, which is consistent with the
0.03 and 0.07 mag computed from the Kurucz model for $T_{\rm eff}=4000$ K, but
lower than the median $([LW2]-[LW5])$ colour derived for stars in the C32
field (Appendix A). The Kurucz model for $T_{\rm eff}=3500$ K as well as the
Fluks et al.\ M5 III model spectrum both yield slightly positive $(K_{\rm
s}-[LW2])$. The red colour of the pulsating AGB star R Cen, $(K_{\rm
s}-[LW2])\sim0.54$ mag, is mostly due to circumstellar emission.

%
%
\begin{table}
\caption[]{For different object classes, $(K_{\rm s}-[LW2])$ colours are
estimated from convolving DENIS and ISO-CAM filter transmission curves with
ISO-SWS spectra (see Yamamura \& de Jong 2000). The rather arbitrary mass-loss
rate $\dot{M}$ subdivisions are indicative of the optical depth of the dusty
circumstellar envelope.}
\begin{tabular}{l|lll}
\hline\hline
Object Class & \multicolumn{3}{c}{$(K_{\rm s}-[LW2])$}                     \\
             & low $\dot{M}$ & moderate $\dot{M}$ & high $\dot{M}$ \\
\hline
AGB (M)      & $0.5\pm0.2$   & $0.9\pm0.2$        & $5.8\pm3.3$    \\
AGB (C)      & $0.3\pm0.2$   & $1.9\pm0.3$        & $6.3\pm1.1$    \\
RSG (M)      & $0.2\pm0.3$   &                    & $3.1\pm0.8$    \\
\hline
\end{tabular}
\end{table}

Combining DENIS K$_{\rm s}$-band magnitudes and ISO-SWS spectra of the M2.5
giant $\beta$ Peg (Glass et al.\ 1999) and the M1 red supergiant (RSG)
Betelgeuse yields negative colours of $(K_{\rm s}-[LW2])=-0.1$ mag for these
stars. For the five RSGs Betelgeuse, Antares, $\mu$ Cep, VX Sgr and NML Cyg
the $(K_{\rm s}-[LW2])$ colour increases monotonically with later spectral
type. The ISO-SWS spectra of a number of AGB stars (Yamamura \& de Jong 2000)
show that $(K_{\rm s}-[LW2])$ colours also become redder with increasing
optical depth of the circumstellar envelope (Table 1).

In conclusion, there is a fair agreement between the observed IR spectra of
red giants that do not have significant circumstellar dust and the Fluks et
al.\ (1994) models which are used here. In particular, $(K_{\rm s}-[LW2])$
colours deduced from observed and model spectra agree for unreddened stars and
are $(K_{\rm s}-[LW2])\sim0$ mag. Further development of the MARCS code (e.g.\
Decin et al.\ 2000) will improve the available model spectra for very cool
stars of type M6 and later, and dynamical models are envisaged to improve the
model spectra for pulsating stars such as the Mira variables.

\subsection{Mid-IR emission from circumstellar dust}

When AGB stars, red supergiants and possibly stars near the tip of the RGB
enter a phase of intense mass loss, they develop dusty circumstellar envelopes
that absorb energetic photons  whose energy is re-emitted at wavelengths of
typically $\lambda\gsim5$ $\mu$m (van Loon 2002 and references therein).
Estimates of the amount of interstellar extinction may therefore contain a
circumstellar extinction component, and the 7 $\mu$m brightness may be
enhanced by circumstellar emission.

Circumstellar dust is easily detected at $\lambda\sim10$ $\mu$m, especially in
the oxygen-rich environments of M-type stars where the silicate dust emission
feature is already strong at moderate mass-loss rates of $\dot{M}\sim10^{-7}$
M$_\odot$ yr$^{-1}$ before the circumstellar envelope becomes optically thick
at $\lambda\sim1$ $\mu$m (van Loon et al.\ 1999). CVF spectra of a sample of
ISOGAL objects (Blommaert et al.\ in preparation) show that at lower mass-loss
rates the ``10 $\mu$m'' feature is probably dominated by emission from
alumina, which is the first dust species to condense. From the $([7]-[15])$
versus $(K_{\rm s}-[7])$ colour-colour diagram (Fig.\ 5), it is clear that
stars with $([7]-[15])\gsim2$ mag must have bright circumstellar emission
(Glass et al.\ 1999; Omont et al.\ 1999; Schultheis et al.\ 2000; Ojha et al.\
2002) and are likely to also experience severe circumstellar extinction. There
are only a few hundred such stars in the sample ($<1$\%), which may be
compared with the 110 ISOGAL counterparts of OH masers found by Ortiz et al.\
(2002) (see Section 4.3). Hence ignoring the separation of the circumstellar
from the interstellar extinction component has negligible impact on our
conclusions.

%
%
\begin{figure}
\centerline{\psfig{figure=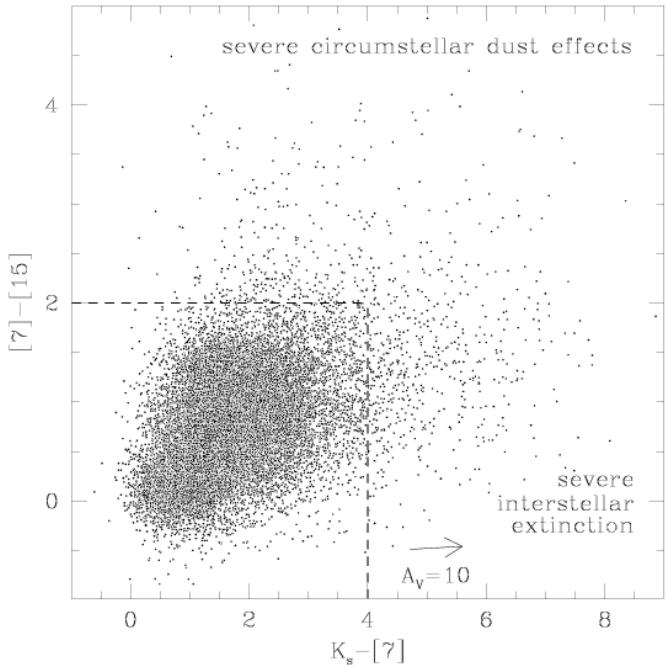,width=88mm}}
\caption[]{IR colour-colour diagram for all our sample stars. Colours
$([7]-[15])\gsim2$ mag are indicative of circumstellar dust emission from
mass-losing giants. Stars without such very red $([7]-[15])$ colours but with
colours $(K_{\rm s}-[7])\gsim4$ mag suffer from extremely severe interstellar
extinction ($A_{\rm V}\sim45$ to 90 mag).}
\end{figure}

Some of the sources with $([7]-[15])>2$ mag may be Young Stellar Objects
(Felli et al.\ 2000, 2002). Most of these will have K$_{\rm s}$-band
counterparts beyond the sensitivity of the DENIS survey, and they will
therefore not appear in Fig.\ 5 nor in the analysis of the ages and
metallicities.

Note also the stars with $([7]-[15])<2$ but $(K_{\rm s}-[7])\gsim4$ mag, which
must experience extremely severe interstellar extinction of $A_{\rm V}\sim45$
to 90 mag (using Mathis' interstellar extinction law: see below). Stars behind
even thicker dust columns are generally beyond the K$_{\rm s}$-band detection
limit. Indeed, at least a few thousand such objects are found within the
ISOGAL data.

\subsection{Isochrones}

The observed stellar populations are compared with theoretical isochrones from
Bertelli et al.\ (1994). The isochrones are paths in $(L,T_{\rm eff})$ space,
and differ according to the age and metallicity of the stellar population.
They are translated into observable colour-colour and colour-magnitude
diagrams in the way as discussed in Section 3.1. The metallicity domain is
restricted to $-2<[M/H]<+0.5$ in order to avoid too large errors due to
extra-polation of the isochrones.

The Bertelli et al.\ stellar evolution computations include OPAL opacities,
Reimers mass loss on the RGB, and convective overshoot. Beyond the onset of
thermal pulses on the AGB, however, the stellar evolution model is replaced by
an analytical population synthesis model to cover the last, less-well
understood, stages of AGB evolution which are characterised by thermal pulses,
radial pulsation, dredge-up and mass loss. The Bertelli et al.\ computations
are very similar to the more recent ones by Girardi et al.\ (2000), despite
some refinements of the physics in the latter, but extend to larger masses
(beyond 7 $M_\odot$) and hence younger ages.

As the vast majority of the ISOGAL sources is expected to be located in the
inner parts of the galactic bulge (van Loon 2001), a uniform distance of 8 kpc
is assumed (Reid 1993; McNamara et al.\ 2000). A range in distances of $\pm1$
kpc is not going to seriously affect the results derived using the isochrones.
The bolometric magnitude of the Sun $M_{{\rm bol},\odot}=4.72$ mag is adopted
(Sterken \& Manfroid 1992).

%
%
\begin{figure*}
\centerline{\psfig{figure=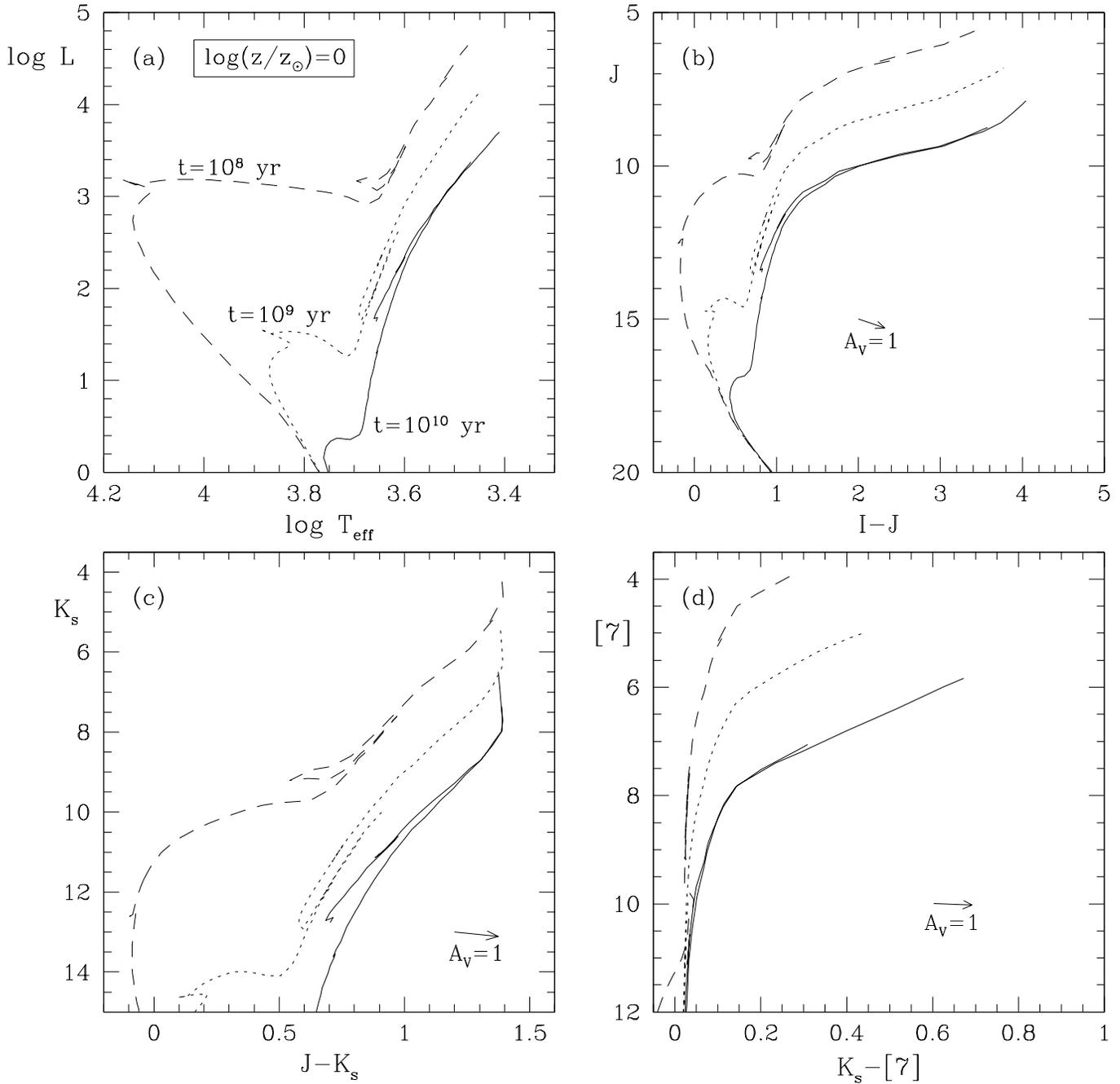,width=180mm}}
\caption[]{Hertzsprung-Russell Diagram from Bertelli et al.\ (1994),
transformed into IR colour-magnitude diagrams. Isochrones are shown for
$\log(z/z_\odot)=0$ and ages of $10^8$ (dashed), $10^9$ (dotted) and $10^{10}$
yr (solid). Extinction (Mathis) is for an M0III star.}
\end{figure*}

%
%
\begin{figure*}
\centerline{\psfig{figure=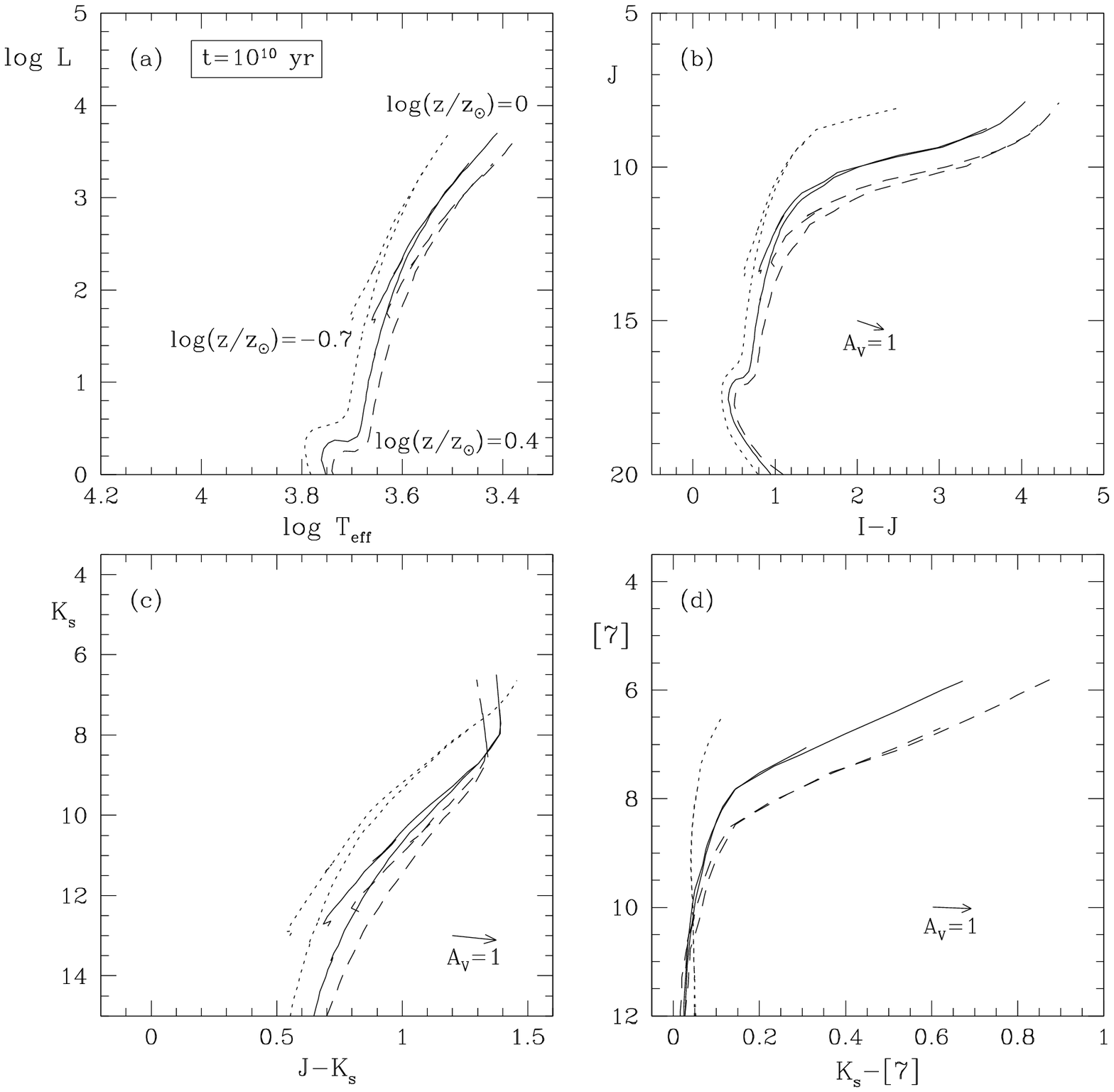,width=180mm}}
\caption[]{Hertzsprung-Russell Diagram from Bertelli et al.\ (1994),
transformed into IR colour-magnitude diagrams. Isochrones are shown for an age
of $10^{10}$ yr and $\log(z/z_\odot)=-0.7$ (dotted), 0 (solid) and $+0.4$
(dashed). Extinction (Mathis) is for an M0III star.}
\end{figure*}

In Figs.\ 6 and 7 some isochrones are plotted in the Hertzsprung-Russell
Diagram and three useful IR colour-magnitude diagrams ordered in increasing
wavelength regime: $J$ versus $(I-J)$, $K_{\rm s}$ versus $(J-K_{\rm s})$ and
$[7]$ versus $(K_{\rm s}-[7])$. The effect of age is illustrated in Fig.\ 6
where three isochrones with solar metallicity are plotted for ages of $10^8$,
$10^9$ and $10^{10}$ yr, whilst in Fig.\ 7 the effect of metallicity is
illustrated by plotting $10^{10}$ yr isochrones for metallicities of
$\log(z/z_\odot)=-0.7$, 0 and $+0.4$.

The effect of metallicity on the $(J-K_{\rm s})$ colours is modest, but the
presence of stars of age $\lsim10^9$ yr will significantly alter the
distribution over $(J-K_{\rm s})$. The brightest J, K$_{\rm s}$-band or 7
$\mu$m magnitudes present are indicative of the age of the youngest stars
present. The $(J-K_{\rm s})$ colour decreases for the most luminous AGB stars
that are very cool and have (super-)solar metallicity, due to selective
molecular absorption. The $(K_{\rm s}-[7])$ colours of the fainter RGB stars
with $[7]\gsim8$ mag are less sensitive to either age or metallicity. The
effect of extinction on the observed colours and magnitudes of the stars is
comparable to the effect of different age and/or metallicity when the
extinction is of order unity.

Interestingly, the sequences in the $(I-K_{\rm s})$ and $(K_{\rm s}-[7])$
versus $(J-K_{\rm s})$ colour-colour diagrams depend on metallicity for the
coolest RGB and most of the AGB stars (Fig.\ 8). This is especially true for
the $(I-K_{\rm s})$ colours, as the I band includes strong molecular
absorption bands of TiO (and VO for spectral types later than M7) that depend
on the metal content of the stellar photosphere. Populations of identical
metallicity but different age fall along the same sequence, but the extension
of the sequence to the reddest IR colours depends on the masses of the most
massive stars present and hence on the age of the population. These cool stars
may experience significant mass loss, affecting their IR colours.

The Bertelli et al.\ isochrones have been used extensively in the past for
deriving ages of clusters from colour-magnitude diagrams (see van Loon et al.\
2001 for an example of the conversion of isochrones to optical and near-IR
colour-magnitude diagrams in order to derive an age for the intermediate-age
stellar cluster HS 327 in the Large Magellanic Cloud). The use of isochrones
in mid-IR colour-magnitude diagrams remains much less explored. Space-borne
mid-IR surveys suffer from poor spatial resolution, hampering the construction
of colour-magnitude diagrams for stellar clusters --- especially down to the
red clump, let alone the Main Sequence. Near-IR photometry has been used in
combination with ISOCAM photometry to construct colour-magnitude diagrams for
the AGB and RGB tip of galactic globular clusters by Ramdani \& Jorissen
(2001) for 47 Tuc, and by Origlia et al.\ (2002) for 47 Tuc, $\omega$ Cen, NGC
6388, M 15 and M 54. These authors made (principal) use of the LW10 filter
($\sim$IRAS 12 $\mu$m filter), although the latter also mentioned the use of
other filters amongst which the ISOGAL survey has (only) the LW6 filter in
common (Origlia et al.\ did not publish their LW6-band photometry). The
general shape of the near/mid-IR colour-magnitude diagrams in these studies
qualitatively agrees with the converted isochrones as shown in Figs.\ 6 \& 7,
with $(J-K)\sim$0.8--1.3 mag and $(K-[12])\sim$0--0.6 mag, while Ramdani \&
Jorissen (2001) show the good agreement between their observed $K$ versus
$(K-[12])$ colour-magnitude diagram and the corresponding isochrone from
Girardi et al.\ (2000).

\subsection{Interstellar extinction}

%
%
\begin{figure}
\centerline{\psfig{figure=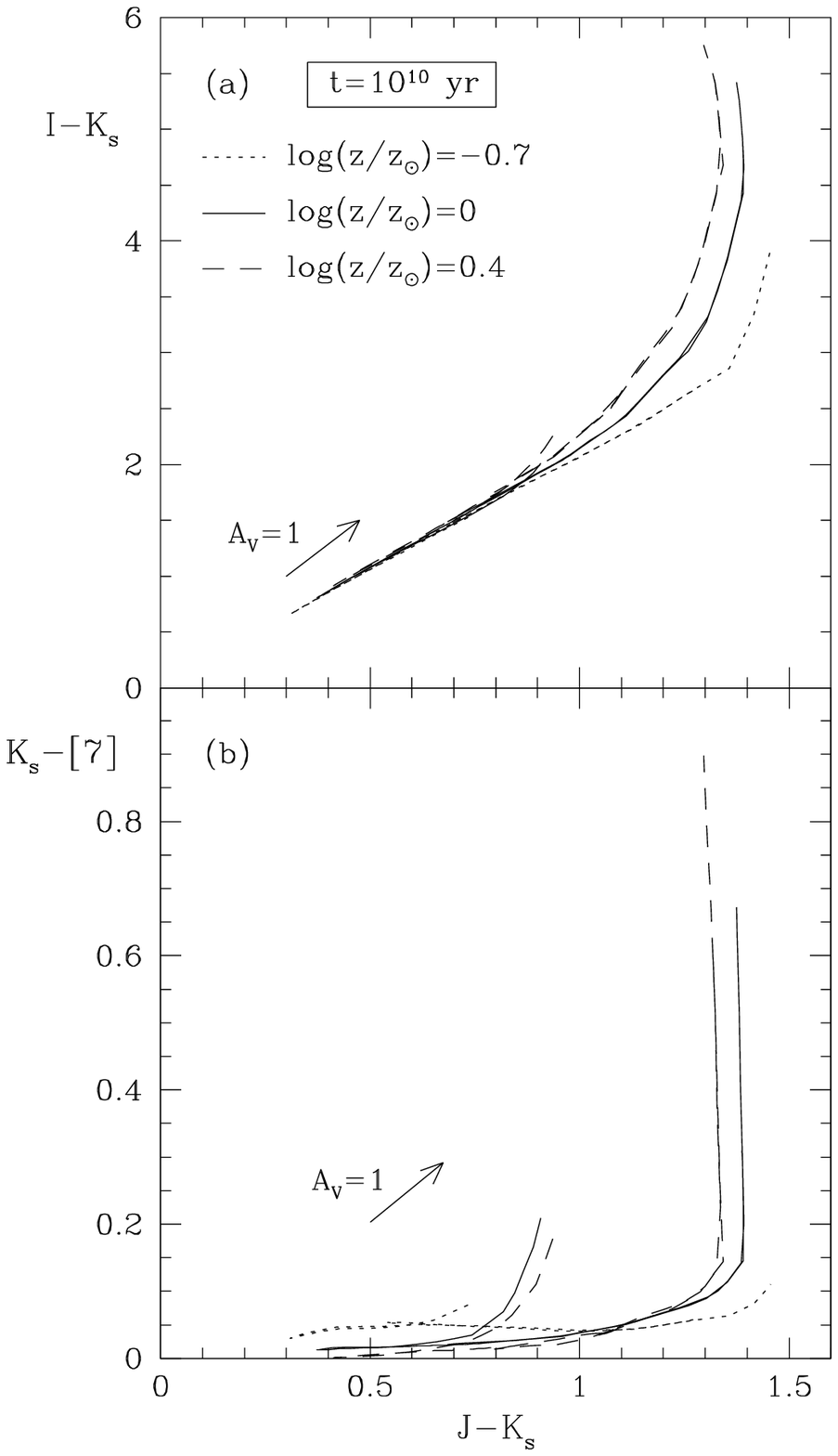,width=88mm}}
\caption[]{IR colour-clour diagrams for isochrones from Bertelli et al.\
(1994), for an age of $10^{10}$ yr and $\log(z/z_\odot)=-0.7$, 0 and $+0.4$.
Extinction is for an M0III star and Mathis' extinction.}
\end{figure}

%
%
\begin{figure}
\centerline{\psfig{figure=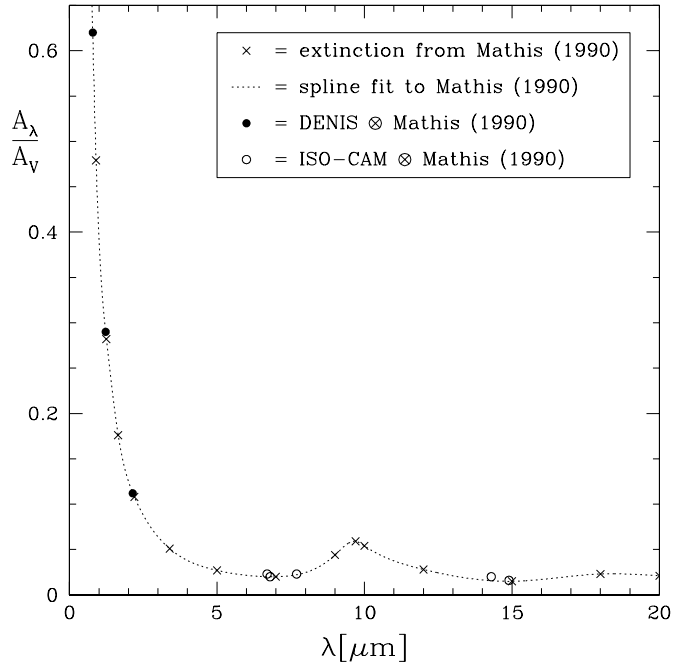,width=88mm}}
\caption[]{Interstellar extinction curve from Mathis (1990), and the
extinction coefficients in the DENIS and ISO-CAM filters for a Vega-like
Kurucz model (spectral type A0V).}
\end{figure}

%
%
\begin{table*}
\caption[]{Extinction coefficients $A_\lambda$ (units of $A_{\rm V}$) in the
DENIS and ISO-CAM filters for different spectral types (see text).}
\begin{tabular}{llllllllll}
\hline\hline
\multicolumn{2}{l}{Spectral Type} &
$A_{\rm I}$                       &
$A_{\rm J}$                       &
$A_{\rm Ks}$                      &
$A_{\rm LW2}$                     &
$A_{\rm LW5}$                     &
$A_{\rm LW6}$                     &
$A_{\rm LW3}$                     &
$A_{\rm LW9}$                     \\
                                  &
                                  &
$\lambda$=0.791$\mu$m             &
$\lambda$=1.228$\mu$m             &
$\lambda$=2.145$\mu$m             &
$\lambda$=6.7$\mu$m               &
$\lambda$=6.8$\mu$m               &
$\lambda$=7.7$\mu$m               &
$\lambda$=14.3$\mu$m              &
$\lambda$=14.9$\mu$m              \\
\hline
A0  & V   & 0.620 & 0.290 & 0.112 & 0.023 & 0.020 & 0.023 & 0.020 & 0.016 \\
M0  & III & 0.611 & 0.287 & 0.112 & 0.023 & 0.020 & 0.023 & 0.020 & 0.016 \\
C5  & III & 0.606 & 0.286 & 0.112 & 0.023 & 0.020 & 0.023 & 0.020 & 0.016 \\
M10 & III & 0.583 & 0.286 & 0.111 & 0.023 & 0.020 & 0.023 & 0.019 & 0.016 \\
\hline
\end{tabular}
\end{table*}

The interstellar extinction curve from Mathis (1990) is folded with the DENIS
and ISO-CAM filter transmission curves and with the Kurucz (1993) Vega-model
spectrum as well as the M-type model spectra from Fluks et al.\ (1994) to
derive extinction coefficients. The extinction curve is plotted in Fig.\ 9 and
the coefficients are tabulated in Table 2.

Interstellar extinction is not always negligible even in the mid-IR, where
$A_{\rm V}=10$ mag still corresponds to at least 0.2 mag extinction. The
extinction coefficients hardly depend on the spectral type of the star, and
the difference only becomes appreciable in the I-band for large $A_{\rm V}$ in
which case most stars will not be detectable in the I-band. Although there is
still some uncertainty in the infrared extinction law, the Mathis law from 0.8
to 7 $\mu$m yields internally consistent results (Section 5.1).

\section{Deriving the metallicity, age and extinction of individual stars}

\subsection{The method}

The method used here essentially fits the Spectral Energy Distribution (SED)
on an individual, star-to-star basis. This differs from fitting isochrones to
a population of stars which assumes uniform extinction, metallicity and age.

Ideally, three variables are solved for: (i) interstellar extinction $A_{\rm
V}$, (ii) metallicity $z$, and (iii) age $t$. We do not (need to) adopt an
age-metallicity relation. Dependent on these variables are the effective
temperature $T_{\rm eff}$, luminosity $L$, and progenitor mass $M$. At least
two observables are used --- K$_{\rm s}$-band and 7 $\mu$m magnitudes --- and
possibly up to four observables are available if the star is also detected in
the I and J-bands. Hence the method cannot always solve for all three
variables, in which case further assumptions are made as described below. The
15 $\mu$m magnitudes are not used because the $([7]-[15])$ colours are more
sensitive to circumstellar emission and photometric scatter than to the
stellar parameters and interstellar extinction. Intrinsic variability
(Schultheis et al.\ 2000; Alard et al.\ 2001) and noise obviously degrade the
extent to which the solution can be constrained.

Schultheis et al.\ (2000) found that $\sim0.2$\% of all DENIS sources are
variable with K$_{\rm s}$-band amplitudes of 0.3 mag or more. Due to
variability of the effective temperature, amplitudes are larger at shorter
wavelengths. At near-IR wavelengths the amplitudes are usually small (few
tenths of a magnitude) and reasonably in phase so that near-IR colours tend to
show little variability. Although mass-losing AGB stars may pulsate with IR
amplitudes as large as 1--2 mag (e.g.\ van Loon et al.\ 1998, 2000, and
references therein), these are very rare objects. The majority of red giants
do not vary much more than the photometric errors (see also Stephens \& Frogel
2002, their Fig.\ 15).

In a colour-magnitude diagram, an object is consistent with one particular
combination of an isochrone and a value for the interstellar extinction. The
solutions from the individual colour-magnitude diagrams are more consistent if
the individual $A_{\rm V}$ values differ less from each other. The best
matching isochrone and interstellar extinction for each star are found by
maximising the parameter $\xi$, which is defined as the quadratic sum of the
inverse differences between the individual $A_{\rm V}$ values and their median
value:
\begin{equation}
\xi =
\sum_{\rm i}{1/\left[\,A_{\rm V,i}-{\rm med}(A_{\rm V,i})+\epsilon\,\right]^2}
\end{equation}
where $\epsilon$ is a small smoothing parameter to avoid singularities and to
decrease the sensitivity to the grid spacing of the isochrones. A value of
$\epsilon=0.02$ mag was found to produce reliable $\xi(z,t)$ maps.

General assumptions include ignoring circumstellar emission (see above) and a
fixed distance $d=8$ kpc. Mathis' interstellar extinction law is assumed to
hold everywhere in the Milky Way, and is applied as explained before: first
for an M0 III spectral type, and in a second iteration for the approximate
solution for the spectral type of the star.

%
%
\begin{figure}
\centerline{\psfig{figure=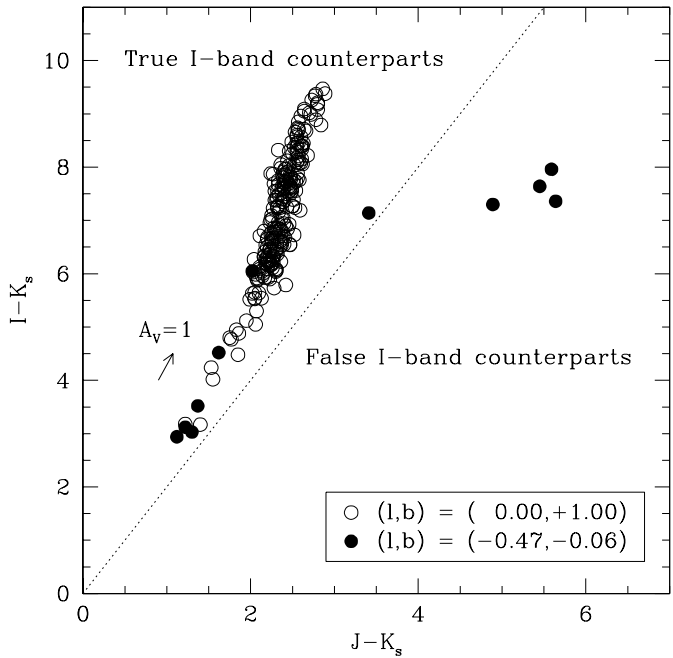,width=88mm}}
\caption[]{$(I-K_{\rm s})$ versus $(J-K_{\rm s})$ colour-colour diagram for
the stars in the moderately extincted C32 field at $(l,b)=(0.00,+1.00)$ and in
the heavily extincted field at $(l,b)=(-0.47,-0.06)$ (see also Fig.\ 1). The
dotted diagonal divides the diagram into a region of realistic stellar colours
(see also Fig.\ 8) and a region with I-band miss-identifications.}
\end{figure}

ISOGAL sources without K$_{\rm s}$-band counterpart are still used in the
analysis of the 7 $\mu$m luminosity distributions. These are assigned the
median value for the interstellar extinction of the nine nearest 7 $\mu$m
sources that have K$_{\rm s}$-band as well as J-band counterparts. In the same
way an estimate for $A_{\rm V}$ is assigned to ISOGAL sources with only a
K$_{\rm s}$-band but no J-band counterpart. For these objects a solution for
the age and metallicity is obtained from the isochrone that best matches the
assigned $A_{\rm V}$ value.

An alleged I-band counterpart is considered a miss-identification and is
subsequently omitted from the analysis if: (i) the ISOGAL source has no J-band
counterpart, or (ii) if $(I-K_{\rm s})<2\times(J-K_{\rm s})-\delta$ mag. The
latter condition becomes clear from Fig.\ 10: realistic stellar colours are
found along a sequence (see also Fig.\ 8), whilst stars with $(I-K_{\rm
s})<2\times(J-K_{\rm s})$ mag appear to have bluer $(I-K_{\rm s})$ colours
than can be explained other than by miss-identification with a (nearly)
coincident but unassociated, relatively bright I-band source. The true I-band
counterpart for such red stars cannot be detected in the DENIS survey. The
$\delta=0.2$ allows for some photometric scatter.

%

Due to the limits on the accuracy of both the isochrones and photometry,
numerical ambiguities, intrinsic photometric variability, the occurrence of
circumstellar emission and extinction, and the presence of foreground objects,
the photometry could not always be satisfactorily matched with any of the
available isochrones. A few thousand such anomalous sources are omitted from
further analysis (see Section 5.3).

%
%
\begin{table}
\caption[]{Source statistics --- excluding 11517 sources without 7 $\mu$m
detection and 3681 anomalous sources.}
\begin{tabular}{llrr}
\hline\hline
Detected             & Not Detected & Number & Percentage \\
\hline
$[7]$                & -            &  35548 & 100\%      \\
$[7]$                & K$_{\rm s}$  &   2108 &   6\%      \\
K$_{\rm s}[7]$       & J            &   6466 &  18\%      \\
JK$_{\rm s}[7]$      & I            &  19346 &  54\%      \\
IJK$_{\rm s}[7]$     & -            &   7628 &  21\%      \\
\hline
\end{tabular}
\end{table}

Table 3 lists some statistics about the number of sources involved, after
correcting for false I-band counterparts and omission of the anomalous
sources. Some fields, especially near the galactic centre, may harbour many
more mid-IR sources than other fields. However, the latter may contain a
significantly higher fraction of sources detected at progressively shorter
(near-IR) wavelengths. Compared to the preliminary photometry, the PSC1.0
catalogue has a higher fraction of associations between mid- and near-IR
sources even amongst the brighter 7 $\mu$m sources.

Different stellar populations may not necessarily be represented in the same
proportions amongst the different groups of stars in Table 3, as the detection
threshold in a certain band may introduce a bias against a certain type of
stellar population. Indeed, the distribution of solutions over metallicity or
age is found to be different for stars with and those without a J-band
counterpart (Section 5): young or metal-poor stars are common amongst the
solutions for stars with a J-band counterpart, but rare amongst those without
a J-band counterpart. This could be the result of a selection mechanism, as
both young and metal-poor stars are brighter and bluer and therefore easier to
detect at progressively shorter wavelengths than old and/or metal-rich stars
which are fainter and redder.

\subsection{A test on artificial data}

%
%
\begin{figure}
\centerline{\psfig{figure=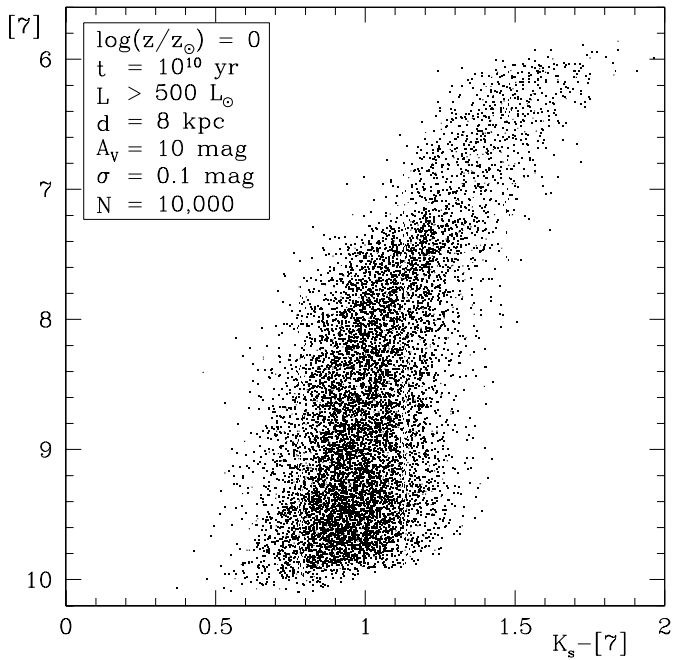,width=88mm}}
\caption[]{$[7]$ versus $(K_{\rm s}-[7])$ colour-magnitude diagram for a
synthetic population, comprising 10,000 stars of 10 Gyr and solar metallicity
that are more luminous than 500 L$_\odot$, placed at a distance of 8 kpc
behind 10 mag of visual extinction. Photometric scatter of $\sigma=0.1$ mag is
added.}
\end{figure}

A synthetic population was created by means of a Monte Carlo simulation,
drawing $N=10^4$ stars from a (Salpeter) initial mass function projected along
the isochrone of Bertelli et al.\ (1994) for an age of $t=10^{10}$ yr and a
metallicity of $[M/H]=0$. Only stars with $L>500$ L$_\odot$ were chosen, which
corresponds to the sensitivity limit of the ISOGAL data at 7 $\mu$m (Section
5.3). Gaussian photometric scatter with $\sigma=0.1$ mag in all photometric
bands was added: this corresponds to a 50\% chance for a deviation in excess
of 0.07 mag. This is probably a conservative limit for the DENIS photometry.
However, as concerns ISOGAL data, such a photometric uncertainty is valid only
for the ISOGAL fields with the best data quality, such as Baade's Window
(Glass et al.\ 1999) and other fields with $|b_{\rm II}|{\gsim}1^\circ$. For
the average ISOGAL data, Schuller et al.\ (2002, Table 12) give
$\sigma\sim0.15$ mag. However, for most of the other fields we consider, close
to the galactic plane, a value $\sigma\sim0.2$ mag would be more appropriate.
The population was placed at a distance $d=8$ kpc behind a visual extinction
of $A_{\rm V}=10$ mag. The photometric scatter causes a small brightness
enhancement at the fainter end of the brightness distribution (Malmquist
bias). The resulting $[7]$ versus $(K_{\rm s}-[7])$ colour-magnitude diagram
is shown in Fig.\ 11.

%
%
\begin{figure}
\centerline{\psfig{figure=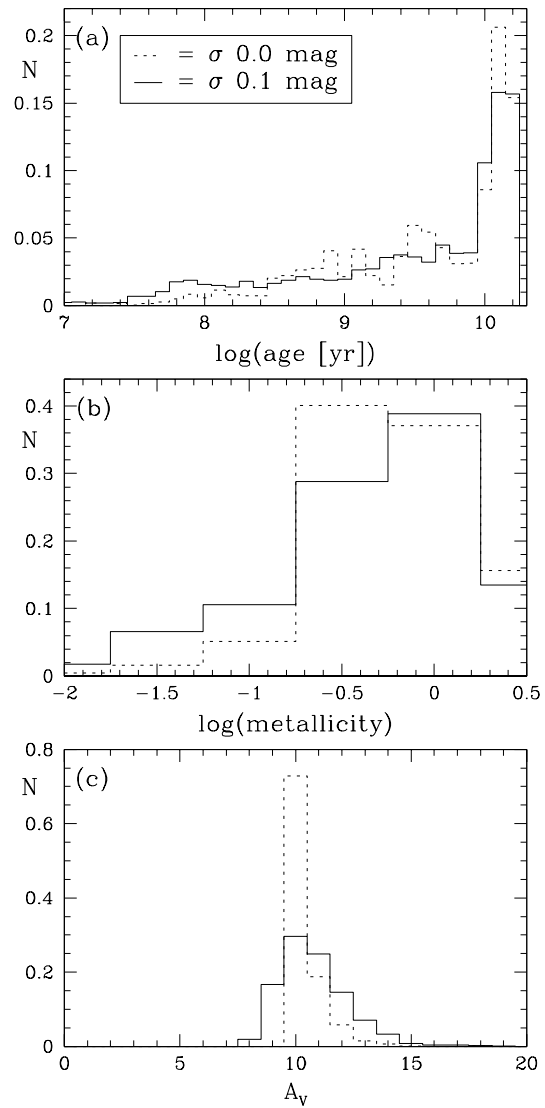,width=70mm}}
\caption[]{Normalised age, metallicity and (visual) extinction distributions
derived for the synthetic population of Fig.\ 11 --- with (solid) and without
(dotted) photometric scatter.}
\end{figure}

For ISOGAL fields with typical extinction of $A_{\rm V}\sim10$ mag the ISOGAL
data is cut at $[7]\sim9$ mag (see Ojha et al.\ 2002). We therefore applied
the same cut to the artificial data, leaving $\sim5700$ sources for further
analysis. The isochrone-fitting method then returned the distributions of age,
metallicity and visual extinction seen in Fig.\ 12. Of the sample with
photometric scatter, roughly 70\% are recovered within a few Gyr of the input
age of $t=10$ Gyr, within ${\Delta}[M/H]\sim0.5$ of the input metallicity of
$[M/H]=0$, and within $\sim2$ mag of the input visual extinction of $A_{\rm
V}=10$ mag. However, the other $\sim30$\% of the sample are distributed down
to ages of $t\sim100$ Myr and metallicities as low as $[M/H]\sim-1.5$.
Photometric scatter as much as $\sigma=0.1$ mag does worsen the spread of
solutions in parameter space, but in particular the determination of the mean
visual extinction is relatively insensitive to photometric scatter. A higher
value of $\sigma$, more appropriate for most of the fields close to the
galactic plane, will slightly worsen the spread of solutions, in particular
for the ages and metallicities. However, the large spread of ages observed in
the results of the simulation even with $\sigma=0$ probably means that the
main problem is the degeneracy of the solution with too poorly constrained
parameters. Relaxing the 7 $\mu$m cut-off or lowering the value of the
extinction does not change the distributions significantly, in particular
those with photometric scatter.

\subsection{Anomalous sources and OH/IR stars}

%
%
\begin{figure}
\centerline{\psfig{figure=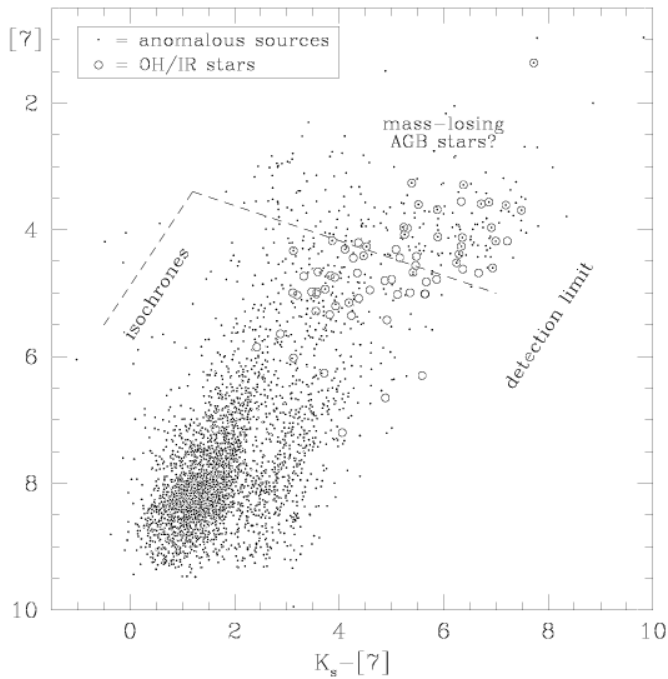,width=88mm}}
\caption[]{$[7]$ versus $(K_{\rm s}-[7])$ colour-magnitude diagram for the
anomalous sources, together with a sample of OH/IR stars.}
\end{figure}

The nature of the anomalous sources --- those that could not be satisfactorily
associated with a single combination of an isochrone and a value for the
interstellar extinction (Section 4.1) --- may be clarified in a $[7]$ versus
$(K_{\rm s}-[7])$ colour-magnitude diagram (Fig.\ 13; dots). Apart from a few
bright and blue sources which are probably in the foreground, the anomalous
sources fall into two groups: those which are too bright in the mid-IR to fit
any isochrone, and those which do occupy the colour-magnitude domain covered
by the isochrones.

In the first group, the reddest bright anomalous sources overlap with the 110
cross-identified OH/IR stars towards the inner galaxy (circles in Fig.\ 13:
Ortiz et al.\ 2002; see also Sevenster et al.\ 1997). Hence we identify the
bright anomalous sources as stars near the tip of their AGB, and severely
reddened by interstellar dust as well as by their own dusty circumstellar
envelope as a result of intense mass loss. Their bright 7 $\mu$m fluxes are
(partly) due to circumstellar dust emission. These stars are also expected to
be variable, which distorts the multi-wavelength photometry as the ISO and
DENIS observations are not simultaneous. Ortiz et al.\ (2002) find that the
OH/IR stars include stars as young as several $10^8$ yr with main-sequence
progenitors of a few M$_\odot$.

The bluer bright anomalous sources could be OH/IR stars too: their relatively
blue colours are due to their location in fields with relatively moderate
interstellar extinction. Judging from the location of the OH/IR stars in fig.\
13, the population of mass-losing AGB stars amongst the anomalous sources
probably includes the anomalous sources down to $[7]\sim6$ mag. Many of these
are not (yet) identified with an OH maser source, either because they have not
been covered by sufficiently sensitive OH surveys or because not all
mass-losing AGB stars exhibit bright OH masers (all of the time). Evidence for
the latter is found, for instance, by Messineo et al.\ (2002) who detect SiO
maser emission from many ISOGAL sources that are not associated with OH maser
emission. The number of objects which are mass-losing AGB star candidates on
the basis of their anomalous colours (a few hundred) is consistent with the
number of objects which are mass-losing AGB star candidates on the basis of
their mid-IR excess emission ($\sim1$\%: Section 3.3).

In the second group, many sources are rather blue and may thus include a
significant fraction of the foreground population. The assumption of a fixed
distance of 8 kpc makes it difficult to reconcile the photometry of foreground
objects with any isochrone/extinction combination, and hence these are
disregarded from further analysis. This, quite conveniently, leaves the
remaining sample of stars (and their subsequent analysis) less affected by the
presence of a foreground population.

\section{The integrated properties of the IR stellar populations in the
central parts of the Milky Way}

\subsection{Extinction}

The elegance of our procedure is that it seeks consistency between the
observed photometry and our current astrophysical understanding. Using the
preliminary photometry, some inconsistencies were found that, at the time,
were remedied by a small modification in the K$_{\rm s}$-band and 7 $\mu$m
extinction coefficients. Using the PSC1.0 photometry, however, the median
values for $A_{\rm V}$ as derived from the different colour-magnitude diagrams
are consistent within $\sim4$\% over the entire sample, confirming the
validity of the Mathis interstellar extinction law. For $\lambda{\gsim}3$
$\mu$m the interstellar extinction law towards the galactic centre may be
higher than the one derived for the solar neighbourhood by a factor two
(Blommaert et al.\ 1998) to four (Lutz 1999; Moneti et al.\ 2001). At
$\lambda\sim3$ $\mu$m the Mathis law lies between those of Rieke et al.\
(1989) and Blommaert et al.\ (1998): the latter combined the van de Hulst
(1949) No.15 law with that from Rieke \& Lebofsky (1985) to include the 10
$\mu$m silicate feature (see also the discussion of the extinction law from
ISOGAL data by Jiang et al.\ 2002). At $\lambda\sim7$ $\mu$m the Mathis law is
similar to Rieke et al.\ but somewhat smaller than Blommaert et al. Doubling
Mathis' extinction coefficient at $\lambda\sim7$ $\mu$m to bring it in line
with Lutz (1999) resulted in a greater discrepancy between the extinction
values derived from the different colour-magnitude diagrams.

%
%
\begin{figure}
\centerline{\psfig{figure=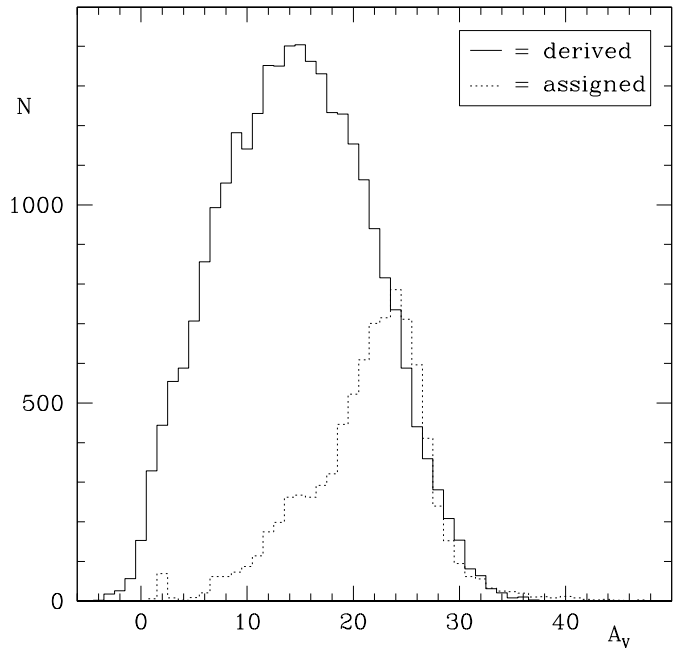,width=88mm}}
\caption[]{Extinction $A_{\rm V}$ derived from IR colour-magnitude diagrams
(solid), and the $A_{\rm V}$ values assigned to stars without J-band
counterparts (dotted).}
\end{figure}

As expected, the extinction is higher, on average, for stars without I-band
counterpart: $A_{\rm V}\sim17$ mag compared to $A_{\rm V}\sim6.7$ mag for the
stars that do have an I-band counterpart. The distribution of extinction
values derived from the colour-magnitude diagrams (Fig.\ 14, solid)
demonstrates the lack of foreground objects amongst the stars that could be
fit by an isochrone/extinction combination: the fraction of sources with
$A_{\rm V}<1$ mag is $\sim1$\% (see also Fig.\ 1), but this does not include
stars in the galactic disk at a distance of a few kpc. The derivation of the
extinction depends on the detection of a J-band counterpart, which in the
DENIS survey becomes problematic when the extinction exceeds $A_{\rm V}\sim20$
mag. Very few J-band counterparts are detected for $A_{\rm V}>30$ mag. Stars
without J-band counterpart have been assigned extinction values derived for
their neighbouring stars, and hence these assigned values are also generally
$A_{\rm V}{\lsim}30$ mag (Fig.\ 14, dotted). Most stars without J-band
counterpart are found in regions of severe extinction, with the extinction
distribution peaking at $A_{\rm V}{\sim}24$ mag compared to $A_{\rm
V}{\sim}15$ mag for the stars with a J-band counterpart.

%
%
\begin{figure}
\centerline{\psfig{figure=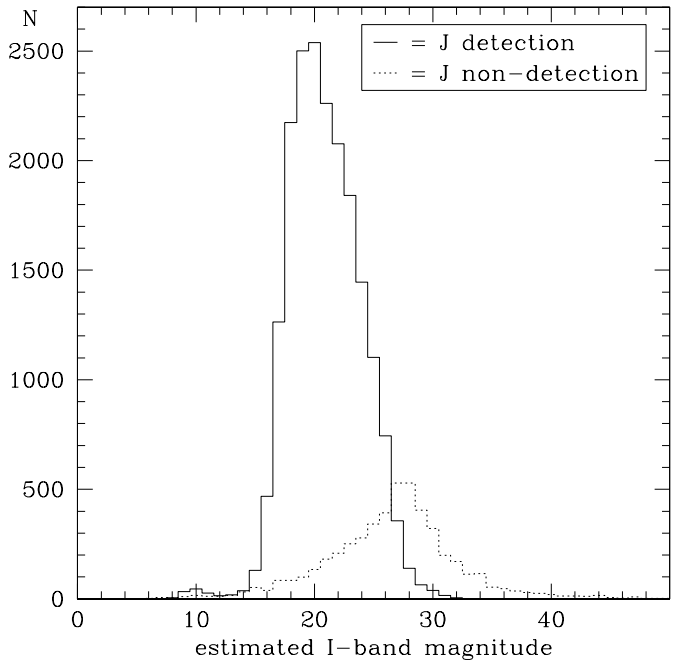,width=88mm}}
\caption[]{I-band estimates for stars without I-band photometry and with
(solid) or without (dotted) J-band photometry.}
\end{figure}

For stars without I-band photometry the I-band magnitude was estimated from
the available photometry and the derived or assigned value for $A_{\rm V}$
(Fig.\ 15). The distributions for the stars with (solid) and without (dotted)
J-band photometry peak at $I=20$ and 28 mag, respectively. Only a negligible
fraction of these I-band estimates have $I<14$ mag; the reasons why these were
missed can be incomplete coverage, blending or saturation. Stars as faint as
$I{\gsim}40$ mag are expected. Clearly, any deeper survey will yield many more
I-band counterparts, but identification may be difficult due to crowding and
angular resolution limitations.

\subsection{Age and metallicity}

%
%
\begin{figure}
\centerline{\psfig{figure=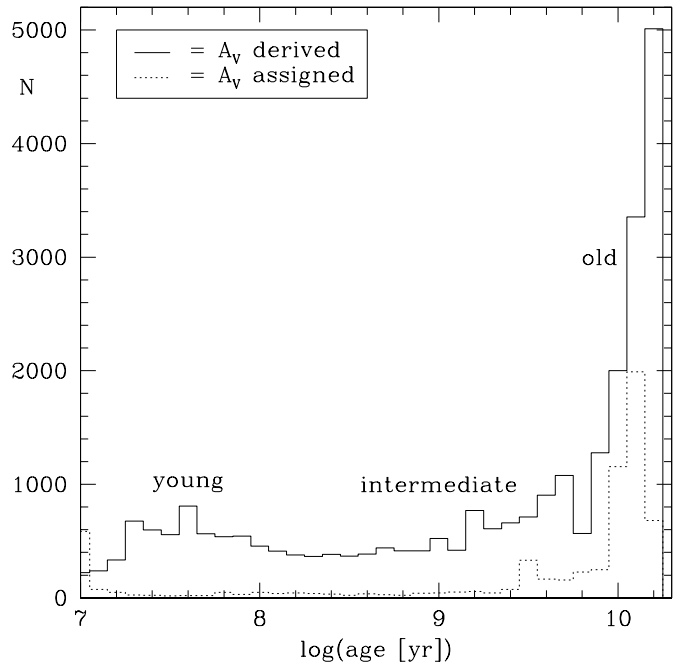,width=88mm}}
\caption[]{Ages of stars derived from the IR colour-magnitude diagrams. Stars
with assigned $A_{\rm V}$ values (dotted) may have been biased against young
stars of relatively early spectral type that are found amongst the less
obscured stars (solid).}
\end{figure}

The distribution over age (Fig.\ 16, solid line) suggests that the inner
galactic bulge consists of three components: old ($t\gsim7$ Gyr),
intermediate-age ($t\sim200$ Myr to 7 Gyr), and young ($t\lsim200$ Myr). The
old population is most populous. The existence of an intermediate-age
population agrees with the presence of mass-losing AGB stars (see Section
4.3). The possible detection of a young population in the inner galactic bulge
is rather intriguing.

%
%
\begin{figure}
\centerline{\psfig{figure=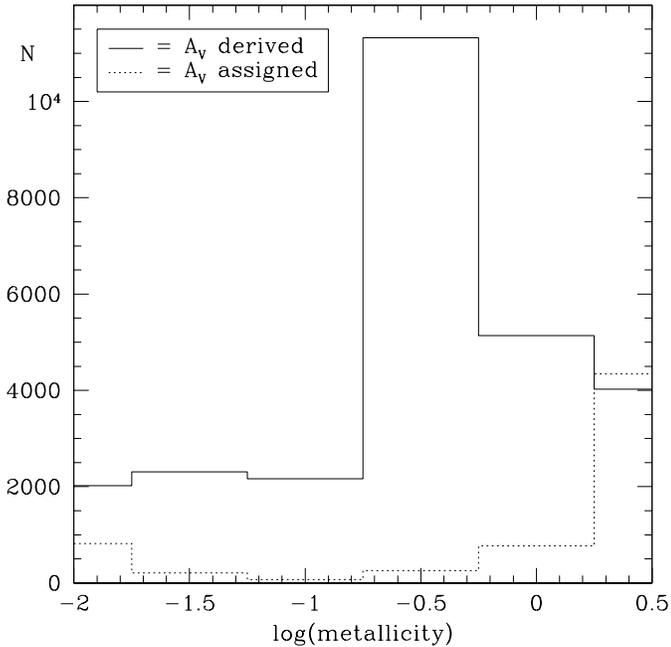,width=88mm}}
\caption[]{Metallicities of stars derived from the IR colour-magnitude
diagrams. Stars with assigned $A_{\rm V}$ values (dotted) may have been biased
against metal-poor stars with relatively blue colours that are found amongst
the less obscured stars (solid).}
\end{figure}

%
%
\begin{figure}
\centerline{\psfig{figure=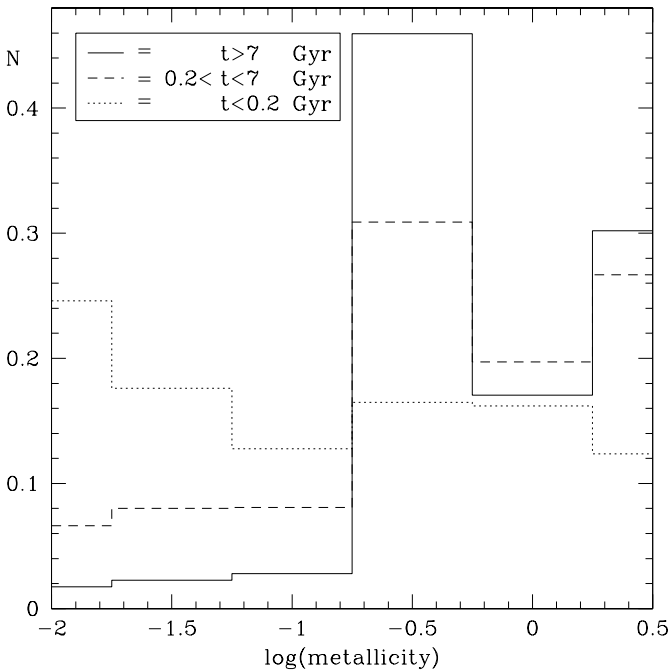,width=88mm}}
\caption[]{The normalised metallicity distribution of stars in different age
groups.}
\end{figure}

The metallicity distribution peaks at $[M/H]\sim-0.5$ (solid histogram in
Fig.\ 17). Stars with metallicities as high as $[M/H]\sim+0.5$ and as low as
$[M/H]\sim-2$ are common, but not dominant. The young stars ($t<200$ Myr)
display a uniform metallicity distribution, whilst the old stars ($t>7$ Gyr)
tend to have higher metallicities (Fig.\ 18). The metallicity distribution of
the old stars may in fact be bimodal, with one component of super-solar
metallicity ($[M/H]\sim+0.5$) and another with sub-solar metallicity
($[M/H]\sim-0.5)$.

The assignment of underestimated $A_{\rm V}$ values for stars without J-band
counterparts may lead to solutions for the age and metallicity that are biased
towards old and metal-rich, because those isochrones tend to correspond to
redder colours whereas young and/or metal-poor stars have generally bluer
colours and would require (even) larger $A_{\rm V}$ values to yield similarly
red colours.

However, it should be stressed that the degeneracy of the solutions and the
photometric scatter tend to distribute the age solutions over the full domain,
as shown by the simulation in Section 4.2. Also, non-recognised foreground
stars may mimic a younger Bulge population (Section 7.1.1). Therefore, if the
ISOGAL results confirm the existence of young and intermediate-age populations
in the inner galactic bulge, their detailed properties remain uncertain.
Because of the age-metallicity entanglement, our conclusions about the
metallicities are equally fragile.

\subsection{Luminosities and temperatures: RGB \& AGB}

%
%
\begin{figure}
\centerline{\psfig{figure=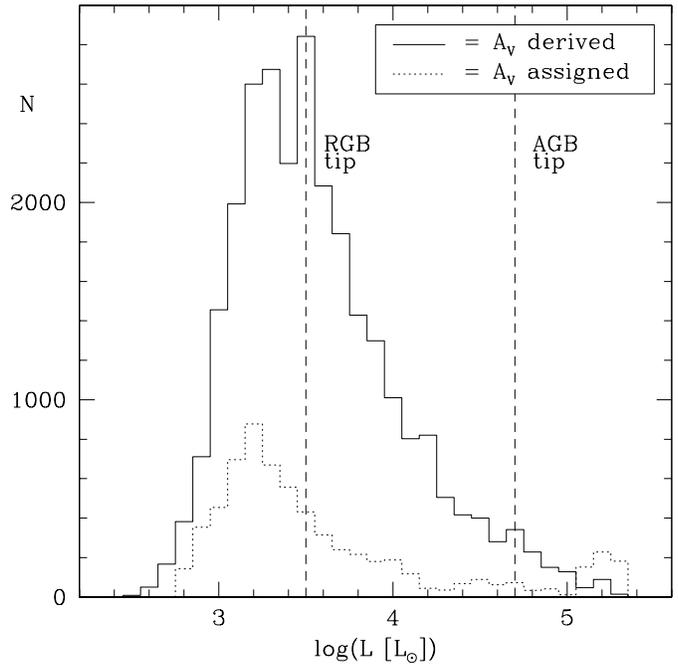,width=88mm}}
\caption[]{The luminosity distribution as derived from the IR colour-magnitude
diagrams.}
\end{figure}

The derived distribution over luminosities extends from $L\sim400$ up to
$\gsim10^5$ L$_\odot$ (Fig.\ 19). The completeness is not fully achieved at
the tip of the RGB at $L_{\rm RGB-tip}\sim3000$ L$_\odot$, and it rapidly
drops below this value. The AGB extends up to $L_{\rm AGB-tip}\sim50,000$
L$_\odot$, although only the most massive AGB progenitor stars ($M\sim5$ to 8
M$_\odot$) are expected to reach such high luminosities. Almost as many RGB
stars as AGB stars are detected. Ultimately, twice as many RGB stars may be
recovered from the ISOGAL data (van Loon \& The ISOGAL Collaboration 2001).
There is no obvious difference in the luminosity distributions of stars with
derived $A_{\rm V}$ and stars with assigned $A_{\rm V}$, except for a small
peak at $L>10^5$ L$_\odot$ for the latter: are these more massive stars in the
central, highly obscured regions of the galaxy? Such conclusion relies mainly
on questionable high values of $T_{\rm eff}\sim2\times10^4$ K (see below).

%
%
\begin{figure}
\centerline{\psfig{figure=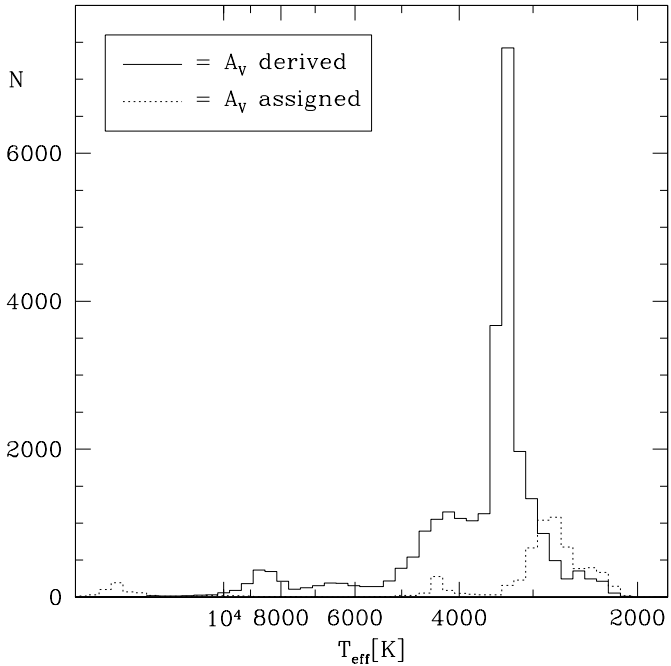,width=88mm}}
\caption[]{The temperature distribution as derived from the IR
colour-magnitude diagrams.}
\end{figure}

The identification of the majority of the ISOGAL sources as red giants is also
confirmed by the distribution over values for $T_{\rm eff}$ (Fig.\ 20), that
sharply peaks at $T_{\rm eff}\sim3300$ K, corresponding to spectral type M6. A
secondary peak in the temperature distribution occurs around $T_{\rm
eff}\sim4200$ K (spectral type late-K). Sadler et al.\ (1996, and references
therein) found $\sim200$ K giants within Baade's Window, compared to a similar
number of M giants. However, Sadler at al.\ stated that ``most K giants with
$V<15.5$ are foreground disk stars'', and fainter K giants are not detectable
by ISOGAL. The temperatures of the stars with assigned values for $A_{\rm V}$
peak at a somewhat lower temperature of $T_{\rm eff}\sim2800$ K (spectral type
M8), but this difference may not be real: the assigned extinction is likely to
be an underestimate of the real extinction. In order to produce the same red
colour as that observed, the inferred temperature would become lower than the
real effective temperature.

%
%
\begin{figure}
\centerline{\psfig{figure=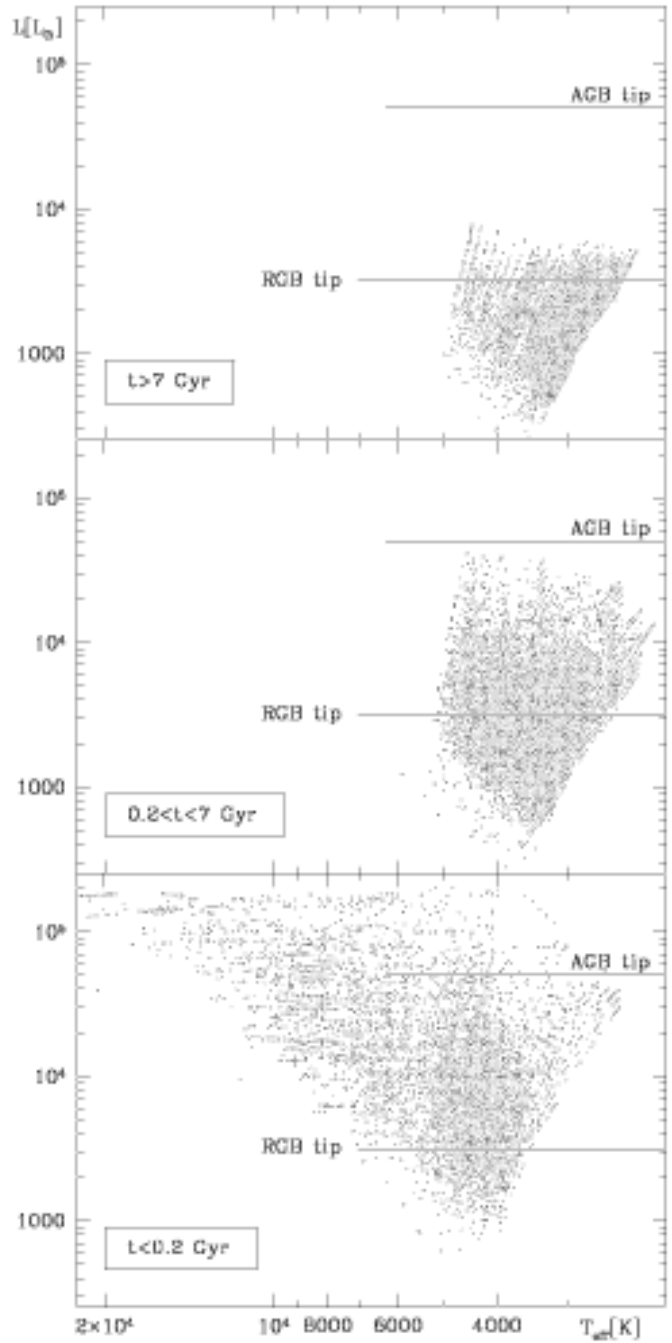,width=88mm}}
\caption[]{The Hertzsprung-Russell Diagram as derived from the IR
colour-magnitude diagrams, separated by age: old ($t>7$ Gyr; top),
intermediate-age ($0.2<t<7$ Gyr; centre) and young ($t<0.2$ Gyr; bottom). The
horizontal lines indicate the maximum luminosity at the tip of the RGB and
AGB, respectively, placed at a distance of 8 kpc.}
\end{figure}

The observed colours and magnitudes of the ISOGAL sources have been tranformed
into their physical properties, of which the temperatures and luminosities can
be plotted in a Hertzsprung-Russell Diagram (Fig.\ 21). The stars have been
separated into three populations of different age. Within each of these
populations there is likely to be a significant spread in metallicity. The old
population ($t>7$ Gyr; Fig.\ 21 top) is traced by RGB stars and faint AGB
stars. The intermediate-age population ($0.2<t<7$ Gyr; Fig.\ 21 centre) is
traced by RGB stars and a well-populated AGB. The young population ($t<0.2$
Gyr; Fig.\ 21 bottom) includes many AGB stars as well as hotter stars: there
is a suggestion of a small population of F-type stars with luminosities
similar to those of Cepheid variables (at $T_{\rm eff}\sim8000$ K and
$L\sim10^{4-5}$ L$_\odot$). Relatively blue and faint stars are not currently
detected in the PSC1.0. For example, a star with $L=1000$ L$_\odot$ and
$T_{\rm eff}=5000$ K is at the limit of detection.

The robustness of these results about the distributions in ages, luminosities
and $T_{\rm eff}$ will be discussed in Section 7.1.

\section{The spatial distribution of the IR stellar populations across the
central parts of the Milky Way}

\subsection{Baade's Window, C32 and (--0.47,--0.06)}

%
%
\begin{figure}
\centerline{\psfig{figure=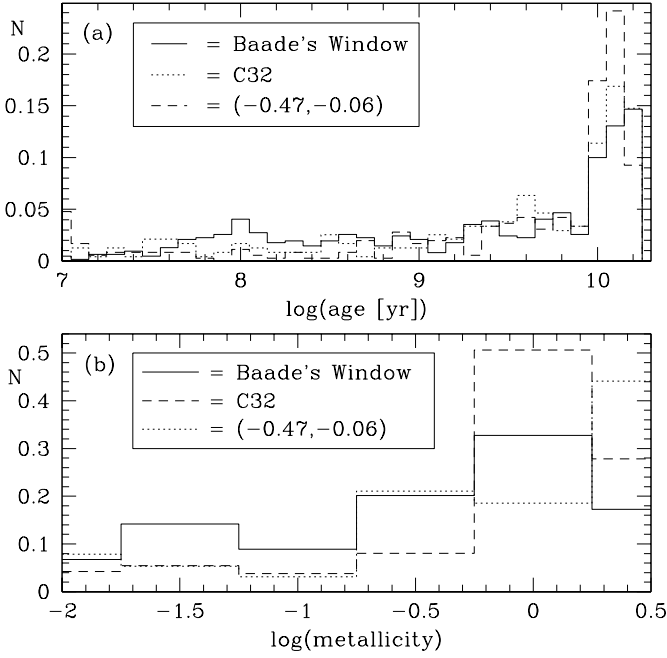,width=88mm}}
\caption[]{The normalised age and metallicity distributions as observed in
Baade's Window (solid), C32 (dashed) and the $(-0.47,-0.06)$ field (dotted).}
\end{figure}

The following three particular fields are investigated first: 633 stars within
a ${\Delta}l_{\rm II}{\times}{\Delta}b_{\rm II}=0.56^\circ\times1.44^\circ$
area in the direction of Baade's Window centred at $(l_{\rm II},b_{\rm
II})=(1.20,-3.23)$, 248 stars within $0.29^\circ\times0.09^\circ$ in the
direction of C32 centred at $(l_{\rm II},b_{\rm II})=(0.00,+1.00)$, and 379
stars within $0.08^\circ\times0.29^\circ$ in the direction of the very dense
$(l_{\rm II},b_{\rm II})=(-0.47,-0.06)$ field. The age and metallicity
distributions are displayed in Fig.\ 22.

All three fields are dominated by old stars. A considerable number of
intermediate-age stars are found, and $\sim20$\% of the stars in Baade's
Window seem to have $t\sim100$ Myr. However, this large number seems
surprising for such a well-studied region and hardly compatible with numerous
previous studies. This finding is certainly contaminated by young star mimics,
for a part to be assessed. They could result either from the instability of
the poorly constrained fitting procedure, enhanced by the uncertainty of the
photometry; or from foreground stars which are difficult to recognise in such
conditions (see Sections 4.2 \& 7.1).

The metallicity distributions are similar for the three fields. They cover a
range in metallicities but are predominantly of solar or slightly super-solar
metallicity. There is a hint that the metallicity increases towards the inner
parts of the Bulge. In Baade's Window, a secondary peak is observed at a low
metallicity of $[M/H]\sim-1.5$, which may be associated with halo stars. It is
interesting in this context to note that Minniti (1996) argues that most of
the RR Lyrae stars found towards the bulge in actual fact belong to the halo.

\subsection{The nucleus, disk and Bulge populations}

The inner galaxy comprises a mix of stellar populations. Four regions are
defined to separate the main components:\\
{\bf (1)} the nucleus within 50 pc of the galactic centre:\\
\hspace*{6mm}$\sqrt{l_{\rm II}^2+b_{\rm II}^2}<0.36^\circ$;\\
{\bf (2)} the central molecular zone (CMZ):\\
\hspace*{6mm}$\sqrt{l_{\rm II}^2+b_{\rm II}^2}>0.36^\circ$, but $|l_{\rm
II}|<2^\circ$ and $|b_{\rm II}|<0.5^\circ$;\\
{\bf (3)} the inner galactic disk:\\
\hspace*{6mm}$2^\circ<|l_{\rm II}|<10^\circ$ and $|b_{\rm
II}|<0.5^\circ$;\\
{\bf (4)} a cross-section through the galactic bulge:\\
\hspace*{6mm}$|l_{\rm II}|<2^\circ$ and $|b_{\rm II}|>0.5^\circ$.\\
Halo stars, due to their low luminosity and low metallicity (blue colours),
are rare amongst the ISOGAL sources.

%
%
\begin{figure}
\centerline{\psfig{figure=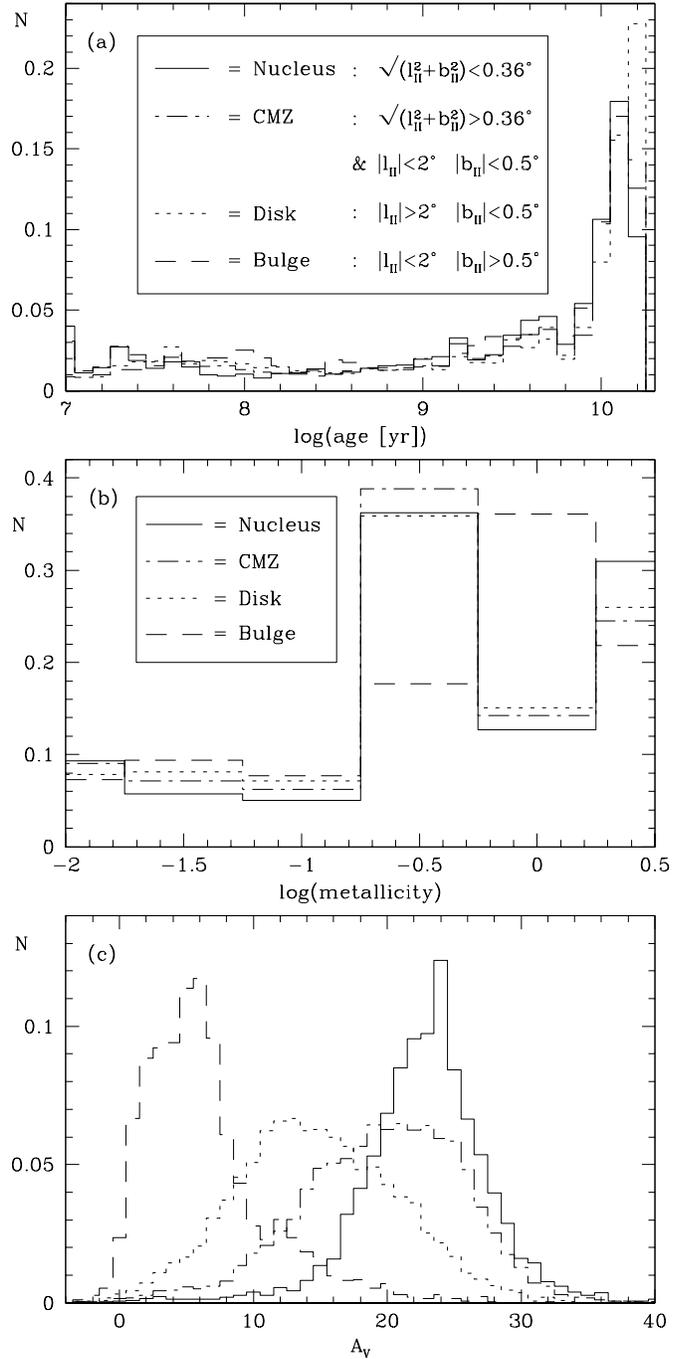,width=88mm}}
\caption[]{The normalised (a) age, (b) metallicity and (c) extinction
distributions for the nucleus (solid), central molecular zone (CMZ:
dot-dashed), disk (dotted) and Bulge (dashed).}
\end{figure}

There is a hint of a trend in the age distributions shifting towards younger
ages when going from the Bulge through the disk and CMZ into the nucleus
(Fig.\ 23a): the old stellar population in the nucleus is a few Gyr younger
than the old stellar population in the Bulge, and the young stellar population
in the nucleus is with a few dozen Myr younger than the young stellar
population in the Bulge ($\sim100$ Myr).

The most significant difference is in the metallicity distribution between the
Bulge and the disk components (Fig.\ 23b): the Bulge clearly peaks around
solar metallicity, whereas the disk components (disk, CMZ and nucleus) seem to
show a bimodal distribution peaking at $[M/H]\sim-0.5$ and $[M/H]\sim+0.5$.
The bimodality seems to be most pronounced in the nucleus, which is also where
the relative number of super-solar metallicity stars is highest.

As expected, the extinction distributions are very different between the
different components (Fig.\ 23c), ranging from $A_{\rm V}\sim5$ mag for the
Bulge to $A_{\rm V}\sim24$ mag for the nulceus. The extinction distributions
through the disk and towards the CMZ are much broader, indicating that much of
the extinction arises from within these components themselves. Considering the
evidence for on-going star formation within the CMZ (Pierce-Price et al.\
2000), it is not surprising that the extinction towards stars in the CMZ is
genrally higher than towards stars in the disk at a few degrees distance from
the CMZ. The fact that the analysis shows clear differences in the extinction
distribution, whilst the age and metallicity distributions are not vastly
different between the various components, demonstrates the ability of the
method to decouple age, metallicity and reddening.

The presence of intermediate-age and young stars in the inner Bulge and the
bimodality of the metallicity distribution in the inner disk and nucleus can
and should be tested by means of near-IR spectroscopy of ISOGAL-selected stars
to determine their metallicities and effective temperatures. At high resolving
powers, kinematic information may help in distinguishing between stars in the
inner Bulge, disk and nucleus from stars in the foreground disk. The study of
mass-losing AGB stars through their pulsation periods and their circumstellar
dust and maser emission has also some diagnostic value with respect to age and
metallicity.

\subsection{Luminosity functions and 3-D structure}

The apparent, dereddened, $[7]_0$ luminosity functions (number of stars per
square degree per magnitude interval: Fig.\ 24) in the inner $\sim2^\circ$
(280 pc) from the galactic centre are skewed towards higher luminosities
compared to regions further away from the galactic centre: the AGB star
population at $[7]_0\lsim7.2$ mag is the dominant contributor in the central
regions whilst the RGB stars are more populous in the off-centre regions. This
is because the interstellar extinction is more severe closer to the galactic
centre, but also because the higher source density and higher background level
closer to the galactic centre result in a brighter limiting magnitude in the
ISOGAL PSC1.0.

%
%
\begin{figure}
\centerline{\psfig{figure=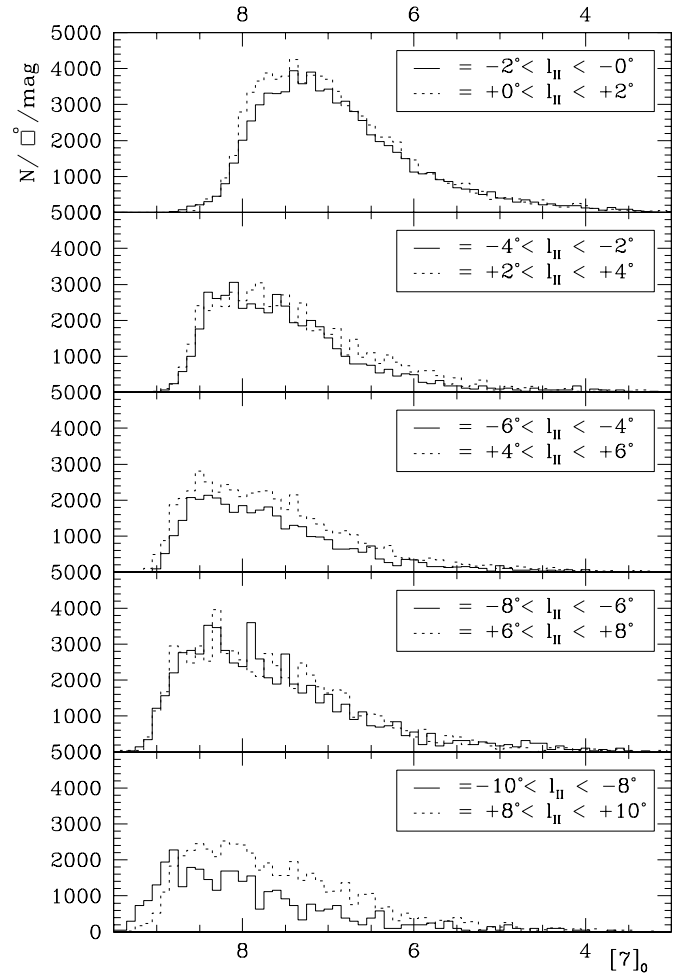,width=88mm}}
\caption[]{The dereddened $[7]$ luminosity functions in $2^\circ$ bins at
either side of the galactic centre, for $|b_{\rm II}|<1^\circ$. An offset is
seen between negative and positive longitudes with $8^\circ{\lsim}|l_{\rm
II}|{\lsim}10^\circ$, indicative of depth effects caused by a bar or spiral
structure.}
\end{figure}

The luminosity functions may probe deviations from the adopted $d=8$ kpc. In
the inner region, at $|l_{\rm II}|\lsim4^\circ$, the luminosity function
(Fig.\ 24) is extremely symmetric around the galaxy's minor axis, indicative
of an azimuthally symmetric spatial distribution of stars. At greater
distances from the galactic centre, however, at $8^\circ<|l_{\rm
II}|<10^\circ$ the luminosity function is brighter at positive than at
negative longitudes. This can be understood if the bulk of the stars at
positive longitudes are located closer to the Earth than those at negative
longitudes.

%
%
\begin{figure}
\centerline{\psfig{figure=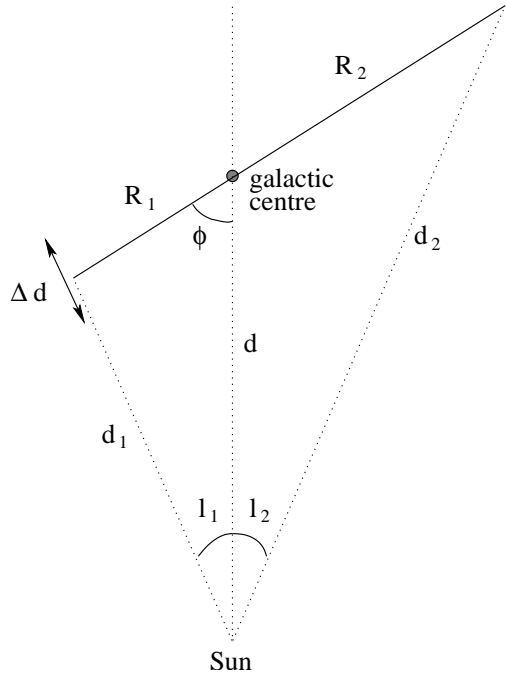,width=66mm}}
\caption[]{Geometry of an inclined bar as seen from the Sun.}
\end{figure}

Suppose that two identical stars, at galactic longitudes $l_1$ and $l_2$, lie
on a line which intersects the galactic centre under an angle $\phi$ --- the
position angle --- with respect to the position of the Sun as seen from the
galactic centre (Fig.\ 25). Then their magnitude difference is:
\begin{equation}
{\Delta}m = -5 \times
\log{\left[\frac{\sin{(\phi+l_1)}}{\sin{(\phi+l_2)}}\right]}
\end{equation}
This also holds for the magnitude difference between the luminosity functions
of stars in a needle thin uniform bar or spiral arm. For a thick structure,
one has to be careful how to normalise the luminosity function before applying
this formula: here, it is normalised to the sampled area.

However, towards the near-part of the bar fewer stars of a particular
luminosity are counted per unit area because the angular separation is larger.
As a result, the bright end of the luminosity function for the near-part of
the bar will be under-populated and the bar angle $\phi$ will be
under-estimated. The vertical scale height of the bar is expected to be small,
though: the sampled areas extend up to $b_{\rm II}=\pm1^\circ$ latitude, where
the stellar density has dropped by an order of magnitude (e.g.\ Alard 2001).
Hence the area effect to the normalisation may not be a quadratic effect
(latitude $\times$ longitude) but rather a linear effect due to the angular
separation in longitude mostly.

The effect of the thickness of the bar in the line-of-sight direction is more
subtle, causing a smearing in the luminosity function. Suppose that one of two
identical stars is located at the near-edge of the bar at a particular
distance from the galactic centre, whilst the other star is located in the
densest part of the bar at the same distance from the galactic centre. Then
their magnitude difference is:
\begin{equation}
{\Delta}m \approx -5 \times \log{\left[ 1 + \frac{{\Delta}d}{d} \right]}
\end{equation}
where $d$ and ${\Delta}d$ are the distance and the difference in the distance
from the Earth, respectively (see Fig.\ 25). Hence the bright end of the
luminosity function will extend to even brighter magnitudes. This effect is
more severe for the near-part of the bar, causing Eq.\ (2) to over-estimate
the bar angle $\phi$.

%
%
\begin{figure}
\centerline{\psfig{figure=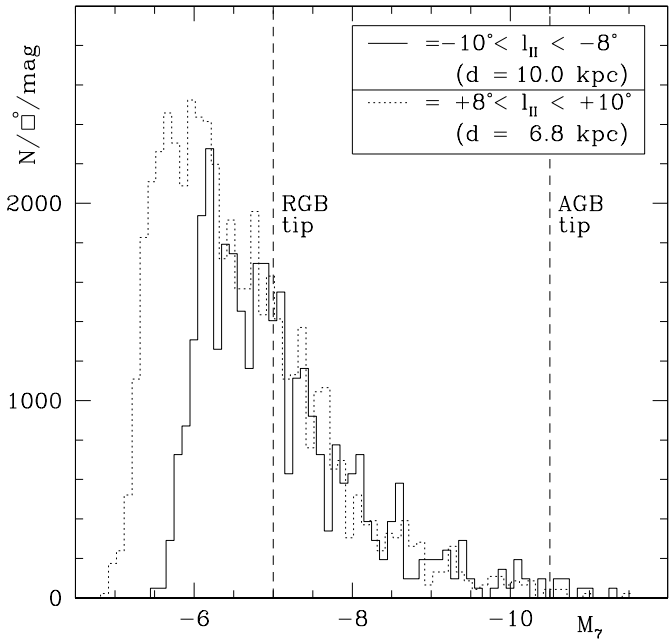,width=88mm}}
\caption[]{Dereddened 7 $\mu$m absolute magnitude distribution functions in
$2^\circ$ bins at $l_{\rm II}\sim\pm9^\circ$ for $|b_{\rm II}|<1^\circ$,
assuming a galactic bar at a position angle $\phi=40^\circ$.}
\end{figure}

If the structure giving rise to the observed depth effects is a bar with the
galactic centre near its point of symmetry, and the net effect of its
thickness is negligible, then according to Eq.\ (2) the luminosity difference
of $\Delta[7]_0\sim0.8$ mag at $l_{\rm II}\sim\pm9^\circ$ corresponds to a
position angle of $\phi\sim40^\circ$. The near end ($l_{\rm II}\sim9^\circ$)
and far end ($l_{\rm II}\sim-9^\circ$) are located at distances of
respectively $d=6.8$ and 10 kpc from the Sun, corresponding to galacto-centric
radii of $R=1.7$ and 2.4 kpc, respectively (Fig.\ 26). Hence there is a clear
distinction between the azimuthally symmetric inner kpc of the galaxy, and the
asymmetric distribution of stars at $R\sim2$ kpc.

\subsection{Interstellar extinction in the galactic plane}

%
%
\begin{figure}
\centerline{\psfig{figure=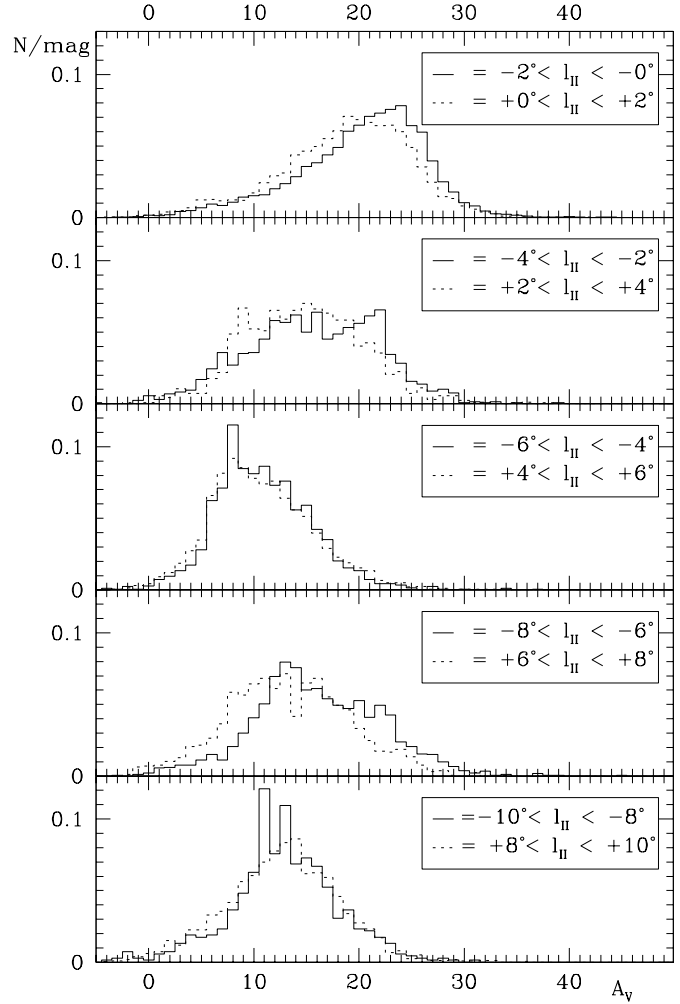,width=88mm}}
\caption[]{The normalised extinction $A_{\rm V}$ distributions in $2^\circ$
bins at several longitudes for $|b_{\rm II}|<1^\circ$.}
\end{figure}

The derived extinction (Fig.\ 27) is typically $A_{\rm V}\sim10$ to 12 mag,
but $\sim$twice as much in the inner $l_{\rm II}\sim\pm2^\circ$. Towards the
inner 40 pc the extinction reaches values of $A_{\rm V}\sim25$ to 40 mag (see
also Cotera et al.\ 2000). Along line-of-sights for $|l_{\rm II}|<5^\circ$ \&
$|b_{\rm II}|<0.5^\circ$ the total extinction through the entire extent of the
Milky Way is estimated from the DIRBE/IRAS dust emission reddening map of
Schlegel et al.\ (1998) to be $A_{\rm V}>90$ mag (Dutra et al.\ 2002). The
results are also in general agreement with Schultheis et al.\ (1999), who
constructed an extinction map from DENIS photometry by fitting a Bertelli
(1994) isochrone of 10 Gyr old and solar metallicity to stars within a
$2^\prime$ sampling window that were assumed to be located at a distance of 8
kpc.

Extinction is known to vary on scales of an arcminute or less (Frogel et al.\
1999), which is one of the reasons for having decided to derive the extinction
for each individual source. The extinction in the direction of $l_{\rm
II}\sim-3^\circ$ and $l_{\rm II}\sim-7^\circ$ is a few mag higher than at
corresponding positive longitudes. These may either be individual columns of
anomalously high extinction, perhaps due to intervening molecular clouds, or
they may be part of a coherent structure such as a spiral arm. Indeed, regions
of relatively low extinction are found at positive longitudes, whilst any
visual light picture of the Milky Way immediately shows obscuration by a
spiral arm at several degrees negative longitude. Within $2^\circ$ from the
galactic centre too, extinction is higher at negative longitudes. This is
where Hennebelle et al.\ (2001) find many dark clouds with mid-IR optical
depths in excess of unity, on scales of typically an arcminute. Launhardt et
al.\ (2002) estimate the volume filling factor of these dense molecular clouds
to be only a few per cent.

\section{Discussion}

\subsection{How robust are the results?}

\subsubsection{Young stars or foreground stars?}

%
%
\begin{figure}
\centerline{\psfig{figure=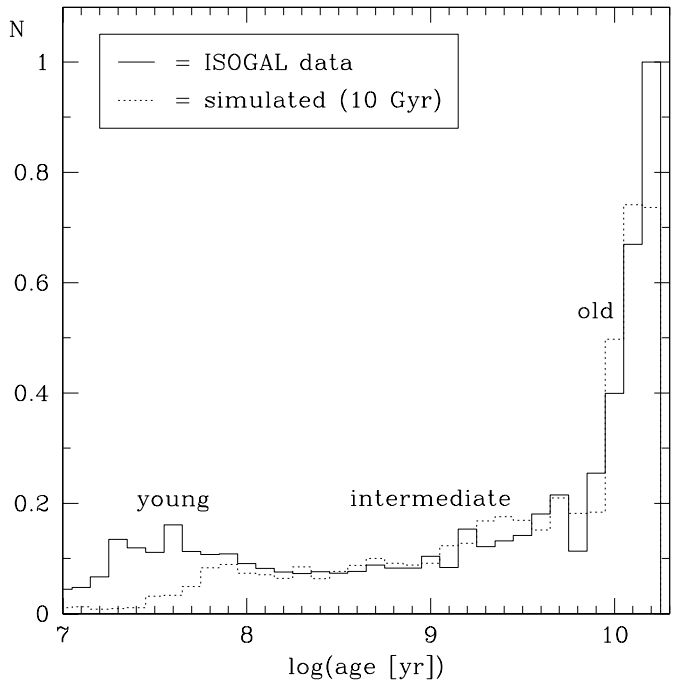,width=88mm}}
\caption[]{Age distribution derived for the ISOGAL data (solid), compared to
the age distribution as simulated by the artificial 10 Gyr-old population with
$\sigma=0.1$ mag (dotted). The young population in the ISOGAL data is not an
artifact.}
\end{figure}

Whether or not of young age, the ``young'' ISOGAL stars represent a population
which is distinct from the old bulge. The simulation of a 10 Gyr-old bulge
population with a photometric scatter of $\sigma=0.1$ mag (Section 4.2) fails
to yield a sufficient percentage of young star mimics (Fig.\ 28). In fact, the
artificial star experiment is a pessimistic simulation of the ISOGAL analysis
in the sense that it over-estimates the population of intermediate-age star
mimics: the simulated age distribution resembles that of the ISOGAL data at
intermediate ages, despite the abundant evidence for the presence of a real
intermediate-age population amongst the ISOGAL sources (Section 4.3; mid-IR
excess from circumstellar dust and OH and SiO maser emission indicate heavy
mass loss attributed to intermediate-age AGB stars) and from other work (see
Section 7.2.2; near-IR surveys by, e.g., Sharples et al.\ 1990). Hence, the
young stars are probably not caused by spurious results due to the method of
analysis.

The failure of the method to isolate the intermediate-age population from the
old and young populations may be due to a combination of selective rejection
and insufficient resolving power. Mass-losing AGB stars of intermediate age
may have been preferentially rejected as a consequence of their variability
and IR excess. Also, the spread of solutions, inherent to the method and the
quality of the data and isochrones, amounts to a spurious intermediate-age
population of $\sim30$\% of the total. The method can therefore only be
expected to resolve intermediate-age populations that constitute at least a
similar fraction of the total.

Feltzing \& Gilmore (2000) argue that a conspiracy of reddening and distance
of foreground stars may mimic a young Bulge population. Foreground stars are
expected to appear not only too young but also generally metal-poor. Thus, the
inclusion of foreground stars in the analysis may cause the spurious detection
of young stars, as well as an upturn in the metallicity distribution for the
young stars at metallicities of $[M/H]<-1$.

OGLE studies of the stars in Baade's Window suggest that a perturbation in the
stellar density and dust extinction is associated with the foreground
Sagittarius arm, at $\sim2$ kpc from the Sun (Paczy\'{n}ski et al.\ 1994).
From a kinematic study in Baade's Window at $b_{\rm II}=-4^\circ$, however,
Sharples et al.\ (1990) found that only 14 out of 239 late-type giants are
foreground stars. Omont et al.\ (1999) estimate that $\sim7$\% of the 7 $\mu$m
sources in the C32 field at $b_{\rm II}=1^\circ$ are foreground objects, but
these do not include foreground stars at a few kpc distance between us and the
galactic centre. Ojha et al.\ (2002) use the simple classical model for the
galactic structure of Wainscoat et al.\ (1992) to estimate that $\sim16$\% of
the ISOGAL mid-IR sources towards $(l_{\rm ii},b_{\rm ii})=(0,\pm2)$ are
foreground stars. A certain number of stars that we have identified as young
Bulge stars may indeed be foreground stars instead.

The a-priori rejection from further analysis of objects believed to lie in the
foreground will necessarily bias any results against the presence of a young
Bulge population. Similar biases are inherent in the selection of M giants for
spectroscopic determination of abundances, in the sense of eliminating
metal-poor stars (Wyse 1999) and the increasingly shorter-lived evolved stages
of massive star evolution. The use of Planetary Nebulae (PNe) as age
determinants (Cuisinier et al.\ 2001) is biased against massive
intermediate-age stars ($M\sim3$ to 8 M$_\odot$) whose PN stage lasts only
very briefly, and against massive stars ($M>8$ M$_\odot$) whose post-RSG phase
lasts even shorter --- if they experience such evolutionary phase at all.

Our analysis excludes objects with photometry that is inconsistent with a star
{\em of any age or metallicity} at a distance of 8 kpc. Hence the resulting
sample is free from any bias against young stars, whilst some fraction of
foreground objects will have been rejected in the process. The distributions
of extinction values are very similar for all age groups (Section 5.1) --- in
particular there is no strong tendency for the young stars to suffer from
especially low or high extinction: the young stars appear to be located at
$d\sim8$ kpc. Also, the young population is traced along its entire
evolutionary path through the HRD (Section 5.3), which is another indication
that the ``young'' stars are indeed of young age and not due to isochrone
mismatch of foreground stars.

The statistical detection of a young population of stars in the inner parts of
the galaxy has thus to be considered, even though some of the individual stars
may have been erroneously classified as young. Similar weaker arguments hold
in favour of the observed bimodal metallicity distribution and the presence of
a significant number of sub-solar metallicity stars. For instance, the
differences in the age and metallicity distributions between the Bulge and
disk/core populations (Section 6.2) cannot be explained by foreground stars or
photometric scatter alone. Also, the simulation experiment (Section 4.2)
produces an artificial tail in the distribution of solutions towards low
metallicities, but not a distinct component around $[M/H]=-1.5$ like that
observed in the Bulge (Fig.\ 22).

\subsubsection{Departures from d=8 kpc}

One may worry about the validity of the fixed distance of $d=8$ kpc that was
assumed in an analysis whose results suggest that in some directions stars are
located at distances that differ from this by $\sim30$\% (Section 6.3).
Suppose that the age and metallicity distributions are identical for the
stellar populations located on either side of (and at similar distance from)
the galactic centre. Then, any spurious effect resulting from a difference in
mean distance will show up as an apparent difference in the derived age and
metallicity distributions. In fact, the age and metallicity distributions for
the stars in the $l_{\rm II}\sim\pm9^\circ$ (and $|b_{\rm II}|<1^\circ$)
directions are indistinguishable (Fig.\ 29), which suggests that the distance
difference has {\em not} seriously affected the derived ages and
metallicities. The distance effect is (even) smaller for the majority of stars
in the sample under study, because they are generally located (much) closer to
the galactic centre where their distances are nearer to the adopted value of
$d=8$ kpc.

%
%
\begin{figure}
\centerline{\psfig{figure=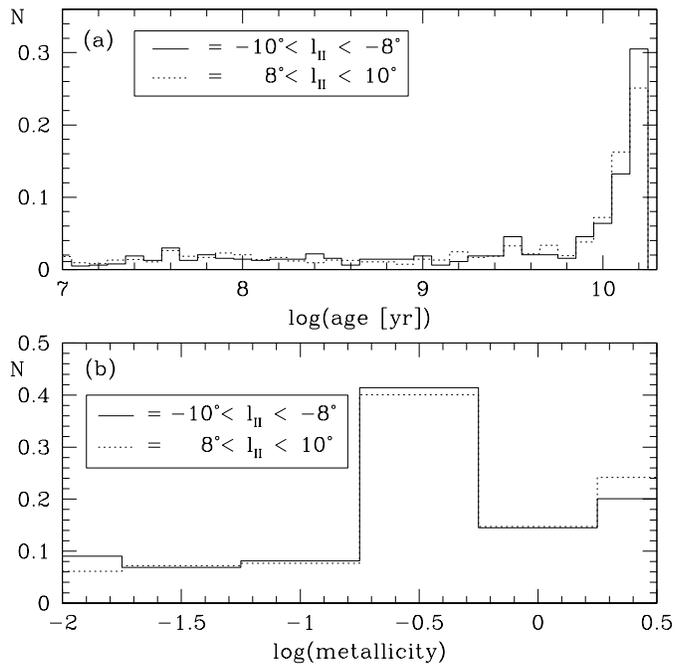,width=88mm}}
\caption[]{The normalised age and metallicity distributions for the inner
galactic plane ($|b_{\rm II}|<1^\circ$) at $l_{\rm II}\sim\pm9^\circ$.}
\end{figure}

The relative insensitivity to distance of the derivation of ages and
metallicities may be explained by the fact that these properties are more
sensitive to colour --- which is distance independent --- than to brightness,
as evidenced by the near-vertical sequences in the IR colour-magnitude
diagrams (Figs.\ 6 \& 7). Changes in distance resulting in changes in the
apparent brightness of order a few tenths of a magnitude do not therefore
strongly affect the derivation of ages and metallicities. This insensitivity
to distance makes it also more difficult to distinguish foreground stars from
Bulge stars, and thus relatively young Bulge stars from foreground mimics.

\subsection{Galaxy formation \& evolution}

\subsubsection{The old population}

The galactic bulge is believed to be old ($t\sim10$ Gyr) and metal-rich
($[M/H]\sim$solar: Rich 2001), indicating that most of it must have formed
early in the history of the Milky Way galaxy on a sufficiently massive scale
to have ensured efficient chemical self-enrichment (Rich \& McWilliam 2000).
The inner Bulge is more complex, however, with the identification of
additional components belonging to the galactic nucleus and disk (see also
Wyse 1999).

The ISOGAL/DENIS data clearly identify the old galactic bulge population as
the main constituent of the inner galaxy. The median age of this population
seems to be somewhat older in the Bulge than in the inner few 100 pc of the
disk. Did the stars in the Bulge form first? Did star formation cease in the
Bulge whilst it continued in the central regions of the galaxy? Did the oldest
stars form in the galactic nucleus but diffuse into the Bulge? Or was it a
combination of the above?

The stars' metallicities, both in the Bulge and inner disk, seem $\sim$solar
on average. This is consistent with the average metallicity of $[M/H]\sim-0.2$
derived from spectroscopy of Bulge K and M giants (McWilliam \& Rich 1994;
Tiede et al.\ 1995; Sadler et al.\ 1996; Ram\'{\i}rez et al.\ 2000b). However,
whereas the metallicity distribution of the Bulge stars is singly peaked at
$[M/H]\sim0$, the metallicity distribution of the disk stars could be bimodal,
with peaks at $[M/H]\sim+0.5$ and $[M/H]\sim-0.5$.

Near the galactic nucleus, the super-solar component seems prominent, but at
larger galacto-centric distances of $\sim1$ kpc the sub-solar component might
dominate (Figs.\ 23 \& 27). This suggests that, besides the more spherical
metal-rich population, a flattened and moderately metal-poor disk population
may be an important component of the inner kpc of the galaxy. A bimodal
metallicity distribution might also explain the properties of OH/IR stars in
the galactic centre, where Blommaert et al.\ (1998) and Wood et al.\ (1998)
argue for a mixture of two populations: stars with fast winds and stars with
slow winds.

%

The relatively high metallicity of the stellar populations in the nucleus and
in the Bulge suggest a link between their formation, with stronger chemical
enrichment of the stars in the galactic nucleus. Perhaps the bulk of the old
stellar population in the galactic nucleus formed after the stars in the
Bulge, whilst chemical enrichment proceeded. An early start to the formation
of the Bulge is inferred from other data (Ortolani et al.\ 1995), whilst an
extended formation period is supported by kinematical data (Zhao et al.\
1994). Early star formation in the Bulge itself, and later kinematical mixing
with newly formed stars from the nucleus naturally reconciles these two
scenarios.

Carraro (1999) suggests a period of quiescence between the formation of the
Bulge around $t\sim13$ Gyr and the formation of the disk at $t\sim9$ Gyr.
Binney et al.\ (2000), however, shows evidence from the solar neighbourhood
that indicates that the formation of the galactic disk must in fact have
started before the end of the great star formation epoch that dominated the
galactic nucleus some 10 Gyr ago. However, the sub-solar metallicity implies
that the infant disk did not experience the intense chemical enrichment that
took place in the galactic nucleus. The galactic disk probably originated with
a metallicity $[M/H]\sim-0.7$ (Rocha-Pinto et al.\ 2000), which could be
confirmed by our observation that the old disk stars seem to have
$[M/H]\sim-0.5$. Note the similarity with the metal-rich component of the
bimodal globular cluster population of the Milky Way (Barmby et al.\ 2000):
did these globular clusters form in the disk? Perhaps the galactic disk formed
from metal-poor material accreted from the galactic halo, possibly mixed with
metal-rich material which was ejected from the galactic nucleus.

\subsubsection{The intermediate-age population}

There is strong evidence from other work that the galactic bulge contains a
population of intermediate-age stars ($t\sim1$ to several Gyr), e.g.\ from the
presence of OH maser sources that represent the evolved stages of stars with
main-sequence masses of $\sim1$ to a few M$_\odot$. OH/IR stars have been
found in the galactic centre (Lindqvist et al.\ 1990, 1991, 1992a,b; Sevenster
et al.\ 1995; Blommaert et al.\ 1998; Wood et al.\ 1998) and to several
hundred pc outwards into the Bulge (Sevenster et al.\ 1997), but the scale
height of stars with ages $t\lsim1$ Gyr is much smaller ($h\sim100$ pc) than
for those with ages $t\gsim5$ Gyr ($h\sim500$ pc) (Sevenster 1999; Frogel
1999a; Frogel et al.\ 1999).

The identification of a few hundred mass-losing AGB stars in the ISOGAL data
confirms the existence of a population of intermediate-age stars. Their
metallicity as well as spatial distribution seems intermediate between those
of the old ``Bulge'' population and those of the younger ``disk'' population.
The intermediate-age component may be a mixture of stars that were formed
during the aftermath of the initial burst of star formation in the galactic
nucleus (see also Sevenster et al.\ 2000), and the products of continuous star
formation in the disk. Migration of stars from the central parts of the galaxy
outward may have resulted in mixing stellar populations throughout the Bulge
and the disk. Such mixing scenario is corroborated by the discovery of
high-speed OH/IR stars in the galactic centre region by van Langevelde et al.\
(1992). The mixing of stellar populations could explain the presence of
metal-rich intermediate-age stars in the Bulge at 600 pc from the galactic
centre (Sharples et al.\ 1990) as well as the presence of old metal-rich stars
in the solar neighbourhood (Feltzing et al.\ 2001).

\subsubsection{The young population}

There is ample evidence for recent and ongoing star formation activity in the
central few hundred pc of the galaxy (Morris \& Serabyn 1996; Morris 2001),
e.g.\ the presence of populous clusters of massive stars (SgrA$\star$: Eckart
et al.\ 1999; Arches \& Quintuplet: Figer et al.\ 2002), very young IR
clusters and molecular clouds that are able to form new massive clusters
(Dutra \& Bica 2001), massive evolved field stars younger than $\sim100$ Myr
(Mezger et al.\ 1999; Philipp et al.\ 1999; L\'{o}pez-Corredoira et al.\
2001a) and massive main-sequence stars (Launhardt et al.\ 2002). It is
believed that this young generation of stars is specific to the galactic
centre region and does not permeate the galactic bulge (Frogel et al.\ 1999).
Therefore, if confirmed, our detection of young stars ($t\sim100$ Myr) in
Baade's Window would be unexpected.

The median age of the young stars in the Bulge ($t\lsim200$ Myr) seems to
decrease towards the galactic centre. This could suggest that the star
formation activity in the galactic nucleus was enhanced $\sim200$ Myr ago and
continued up to the present day. Serabyn \& Morris (1996) also argue that star
formation in the inner 200 pc of the Bulge has continued over the lifetime of
the galaxy (Gilmore 2001), shaping the observed stellar density cusp in the
central 200 pc of the galaxy (Becklin \& Neugebauer 1968). The young stars at
a few hundred pc out of the galactic plane may have formed closer to the
galactic centre and may then have migrated towards higher latitudes as a
result of scatterings off giant molecular clouds in the central molecular zone
(Kim \& Morris 2001; Pierce-Price et al.\ 2000) or heating of the disk by a
merger event (Wyse 2000) on a timescale of $\sim10^8$ yr.

Given the intense chemical enrichment in the central regions of the Milky Way
galaxy during its early evolution, the wide range in metallicities of the
young ISOGAL/DENIS objects, including many stars of sub-solar metallicity, as
well as the spectroscopically determined solar metallicity of cool
(super)giants in the galactic centre (Carr et al.\ 2000; Ram\'{\i}rez et al.\
2000a) is at first puzzling. This suggests that the metal-rich gas must have
been diluted with metal-poorer gas before the more recent star formation took
place. Such relatively metal-poor gas must originate from outside the central
regions of the Milky Way (CMZ and nucleus). We could speculate that the onset
of the star formation event 200 Myr ago might have been triggered by the
infall of metal-poor gas from clouds in the galactic halo (see Richter et al.\
2001), a passing dwarf galaxy or a minor merger (see also Frogel 1999b). This
may be a generic scenario to explain nuclear star forming activity in most
nearby galaxies.

In conclusion, the presence of a young stellar population in ISOGAL data is
solidly confirmed. However, more work is needed to fully assess its detailed
properties and its importance by well confirming the elimination of spurious
mimics of young stars resulting from the degeneracy of the stellar parameter
determination and from foreground sources. The case of the most luminous
sources found in the HR diagram of young stars (Fig.\ 21) should be addressed
with priority. In particular, the number and the distribution of very luminous
intrinsically blue sources need to be confirmed.

\subsection{The bar}

The dereddened 7 $\mu$m luminosity distributions across the galactic plane
indicate that stars at $l_{\rm II}\sim+9^\circ$ are generally closer to Earth
than stars at $l_{\rm II}\sim-9^\circ$. This can be understood in terms of an
elongated Bulge or the presence of a bar with position angle
$\phi\sim40^\circ$. Gas dynamical models for the inner galaxy require a bar
(Blitz et al.\ 1993). There is growing observational evidence to support the
presence of a bar with a semi-major axis of $R\sim3$ to 4 kpc. The position
angle that we derive is similar to what is inferred for the bar at greater
distances from the galactic centre --- typically $\phi\sim20^\circ$ (Stanek et
al.\ 1997; Englmaier 2000; Gerhard 2001) or $\phi\sim40^\circ$ (Deguchi et
al.\ 1998; Sevenster et al.\ 1999; L\'{o}pez-Corredoira et al.\ 2001b).

Unavane \& Gilmore (1998) argue for the presence of a bar-like structure at
$l_{\rm II}\sim\pm4^\circ$ with a position angle $\phi\sim20^\circ$ but with
the near-side at negative longitudes, which agrees with the detection by Alard
(2001) of a bar with a similar orientation within the central degree of the
galaxy. However, the analysis of the ISOGAL/DENIS PSC1.0 data suggest that,
within $\sim1$ kpc of the galactic centre, the Bulge and inner disk are
azimuthally symmetric. We do observe a small asymmetry in extinction, though,
which may mimic differential depth effects.

The symmetry and observed lack of organised motion in the inner Bulge
(Zijlstra et al.\ 1997; Sevenster et al.\ 2000) suggest that the asymmetry
observed further out is a disk phenomenon, rather than a tri-axial Bulge. The
bar is detected at (and beyond) the inner Lindblad resonance, which is located
at $R\sim1$ to 1.5 kpc from the galactic centre (Sevenster 1999; see also
L\'{o}pez-Corredoira et al.\ 2001b).

\section{Summary}

The combined ISOGAL mid-IR survey and DENIS near-IR survey have for the first
time provided a large sample of stars in the innermost parts of the galactic
bulge and disk. Extinction, age and metallicity distributions are derived by
comparison with isochrones. The main results are:\\
{\bf (1)} Stars are detected down to luminosities below the tip of the RGB,
with the fraction of RGB stars being typically $\sim$50\% in all but the most
obscured sightlines;\\
{\bf (2)} The galactic bulge, disk and nucleus are dominated by an old ($t>7$
Gyr) population. There is a hint that the Bulge is slightly older than both
disk and nucleus;\\
{\bf (3)} The average metallicity of the old Bulge stars is $\sim$solar. The
metallicity distribution of the old stars in the disk and nucleus could be
bimodal: a super-solar metallicity component may be linked to the formation of
the Bulge, and a sub-solar metallicity component is identified with the
disk;\\
{\bf (4)} An intermediate-age (200 Myr $<t<7$ Gyr) population is detected in
the inner galactic bulge, disk and nucleus. Some similarity in metallicities
with the old population suggests that the formation of the intermediate-age
stars is linked to the formation of the old population;\\
{\bf (5)} A young ($t<200$ Myr) population might have been detected in the
inner galactic bulge, disk and nucleus. There is a hint that the young stars
are older in the Bulge compared to those in the disk and nucleus. The broad,
largely sub-solar metallicity distribution of the young stars suggests that
their formation history differs from the older generations;\\
{\bf (6)} The galactic bar is detected at a galacto-centric radius of
$R\gsim1$ kpc. The position angle of $\phi\sim40^\circ$ agrees with the
orientation of the bar at larger galacto-centric distances ($R\sim3$ kpc). The
inner few hundred pc of the galactic bulge/disk appears azimuthally symmetric
in terms of the apparent, extinction-corrected luminosity distribution, but
slightly asymmetric in terms of the amount of extinction.

We interpret these results as evidence for a lengthy and dynamical formation
history of the galactic nucleus and Bulge. Started $\gsim10$ Gyr ago in the
Bulge and nucleus with a high rate of star formation and chemical enrichment,
it continued for several more Gyr in the nucleus: in time, the star formation
in the Bulge has become more centrally concentrated (Frogel 1999a,b). After
formation in the nucleus, stars may have subsequently migrated and mixed into
the Bulge, either by diffusion or by scattering off molecular clouds. The
formation of the disk must have been distinct from that of the nucleus and
Bulge. Star formation activity in the inner few kpc of the disk may have been
sustained by inflow of gas from the galactic halo.

\section*{Acknowledgments}
Jacco wishes to thank everyone at the Institute of Astronomy in Cambridge, UK,
for a wonderful time as a postdoc. This work was carried out in the context of
EARA, the European Association for Research in Astronomy. Based on
observations with ISO, an ESA project with instruments funded by ESA Member
States (especially the PI countries: France, Germany, the Netherlands and the
United Kingdom) and with the participation of ISAS and NASA. Based on
observations collected at the European Southern Observatory, La Silla Chile.
The DENIS project is supported in France by the Institut National des Sciences
de l'Univers, the Education Ministry and the Centre National de Recherche
Scientifique, in Germany by the State of Baden-W\"{u}rtemberg, in Spain by the
DGICYT, in Italy by the Consiglio Nazionale delle Ricerche, in Austria by the
Fonds zur F\"{u}rderung der wissenschaftlichen Forschung und Bundesministerium
f\"{u}r Wissenschaft und Forschung. We thank Harm Habing for his comments on
an earlier version of the manuscript, and the (anonymous) referee for her/his
constructive remarks that helped improve the paper. As Joaninhas s\~{a}o os
animais mais fofinhos do mundo inteiro.

\appendix

\section{ISO data homogenisation}

For different fields, the ISO-CAM observations were done through different
filters and with different pixel sizes. These photometric data can only be
merged if we know the transformation formulae that relate photometry obtained
with each filter and pixel size combination. Field C32 at $(l_{\rm II},b_{\rm
II})=(0.00,+1.00)$ has been observed through all possible filter and pixel
size combinations, and therefore it serves as the principal calibration field
for the ISOGAL database.

%
%
\begin{figure}
\centerline{\psfig{figure=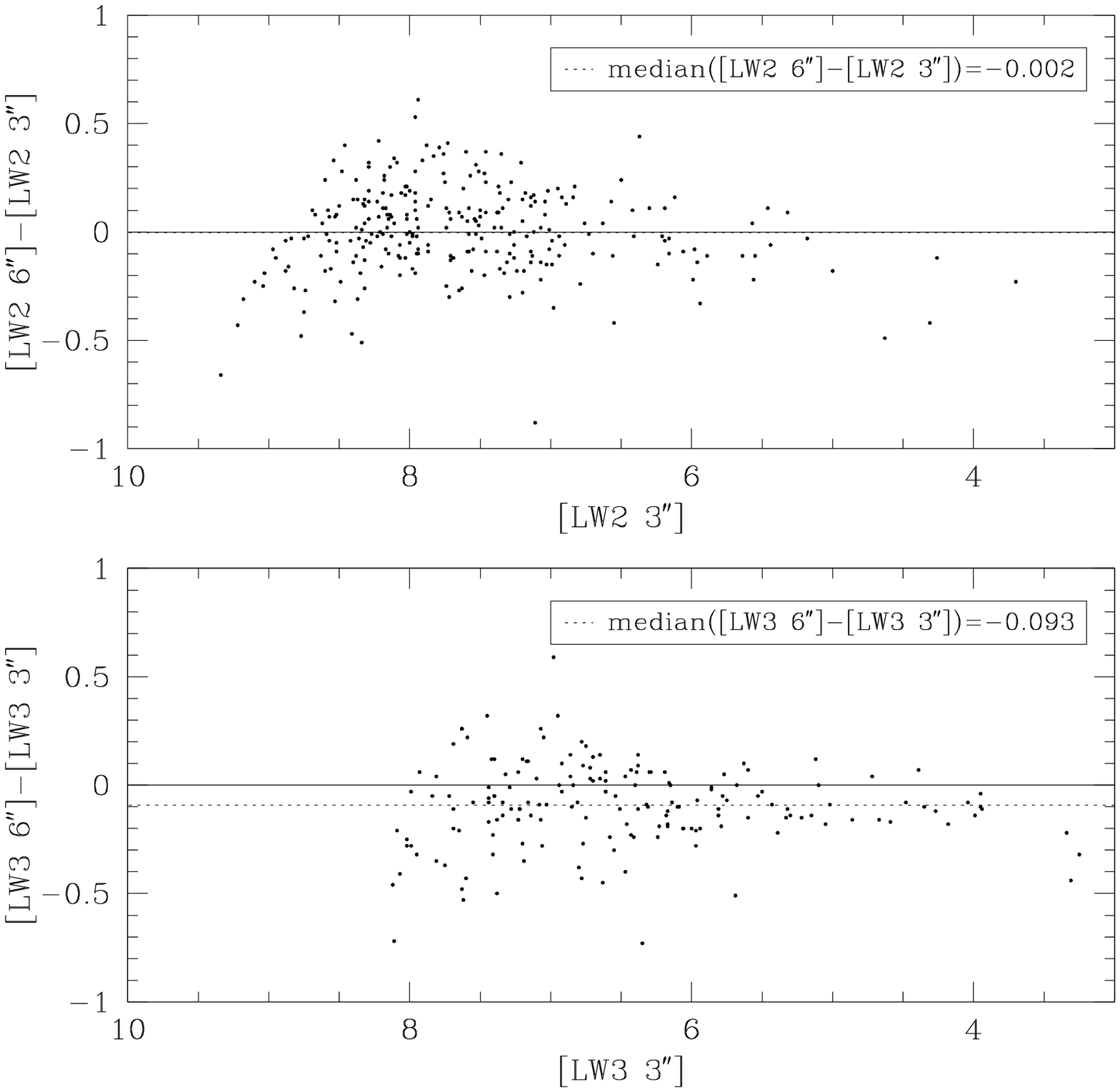,width=88mm}}
\caption[]{Comparison of the photometry of the stars in the C32 field using
different pixel sizes: $3^{\prime\prime}$ and $6^{\prime\prime}$.}
\end{figure}

Photometry at 7 $\mu$m (using the broadband LW2 filter) hardly depends on the
pixel size, at least for the moderately crowded field C32 (Fig.\ A1). However,
at 15 $\mu$m (using the broadband LW3 filter) fluxes of point sources obtained
using a $6^{\prime\prime}$ pixel size are systematically brighter by $\sim0.1$
mag than those obtained using the smaller $3^{\prime\prime}$ pixel size (Fig.\
A1). This may be due to more severe crowding at 15 $\mu$m because of the
larger PSF size at longer wavelengths.

%
%
\begin{figure*}
\centerline{\psfig{figure=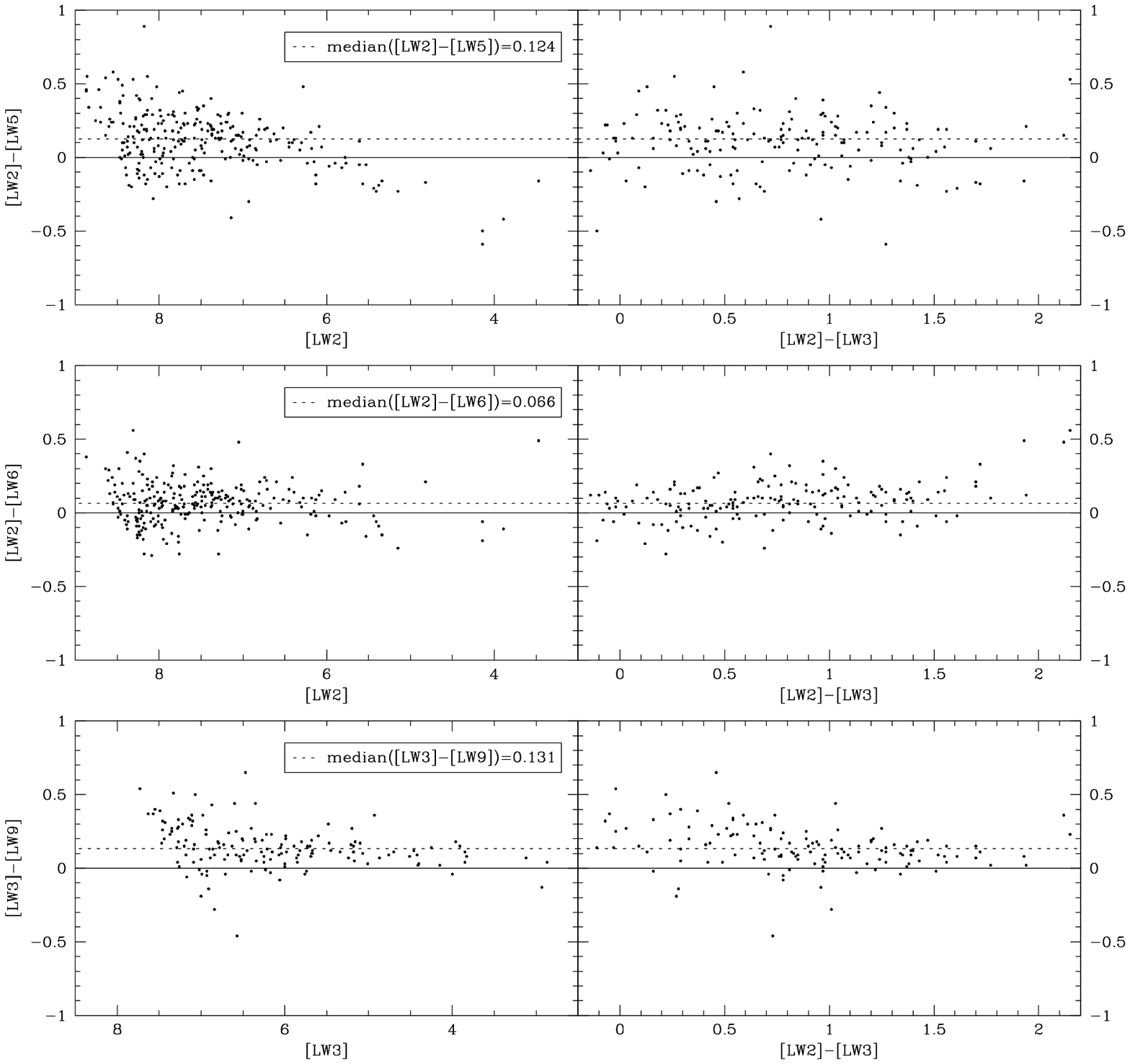,width=180mm}}
\caption[]{Comparison of the photometry of the stars in the C32 field as
obtained through the following different ISO-CAM filters: LW2, 5 and 6 around
a wavelength of 7 $\mu$m, and LW3 and 9 around a wavelength of 15 $\mu$m.}
\end{figure*}

Differences in magnitudes obtained through filters with different passbands
can arise if the spectral slope deviates from that of the Vega model used to
set the zero points of the ISO-CAM photometry. Reasons for this to happen
include differences in $T_{\rm eff}$ (the Rayleigh-Jeans law is not a good
approximation at 7 $\mu$m for very cool stars), severe interstellar
extinction, circumstellar extinction and emission, and the presence of
photospheric molecular absorption bands. These effects tend to make a star
appear redder than the Vega model, except possibly for the absorption bands.

The LW2 and LW5 filters are nearly concentric, yet the photometry for stars in
the C32 field shows a systematic $([LW2]-[LW5])\sim0.12$ mag (Fig.\ A2). The
likely reason for this is that the narrowband LW5 filter largely covers
photospheric continuum, whilst the broadband LW2 filter includes strong
molecular absorption bands around 4 and 8 $\mu$m. As the LW6 filter includes
the 4 $\mu$m but not the 8 $\mu$m absorption band, the fact that the stars
have systematically $([LW2]-[LW6])\sim0.07$ mag (Fig.\ A2) suggests that the
contribution of the 8 $\mu$m absorption band is, roughly, as much as that of
the 4 $\mu$m band. The $([LW2]-[LW5])$ colour is independent of the
$([LW2]-[LW3])$ or $([7]-[15])$ colour, suggesting that the strength of the
molecular absorption does not depend on the IR excess due to emission from
circumstellar dust. This might not be strictly true for the $([LW2]-[LW6])$
colour, and the LW6 filter might in fact include some of the IR excess.

The LW3 and LW9 filters are nearly concentric, but the broadband LW3 filter
may include some of the circumstellar silicate dust emission feature that
peaks at $\sim10$ $\mu$m, which probably competes with molecular absorption in
determining the $([LW3]-[LW9])$ colour. The stars in the C32 field have
$([LW3]-[LW9])\sim0.13$ mag, becoming bluer when circumstellar emission
$([LW2]-[LW3])$ increases (Fig.\ A2).

\label{lastpage}
\end{document}